# Resolving Structural Transitions in Lanthanide High-Entropy Oxides

Billy E. Yang,[1] Saeed S. I. Almishal,[1] Sai Venkata Gayathri Ayyagari,[1] Mary Kathleen Caucci,[2] Gerald Bejger,[3] Christina M. Rost,[3] Nasim Alem,[1] Susan B. Sinnott,[1,2,4] Jon-Paul Maria[1]

[1]Department of Materials Science and Engineering, The Pennsylvania State University, University Park, PA 16802, USA

[2]Department of Chemistry, The Pennsylvania State University, University Park, PA 16802, USA

[3]Department of Materials Science and Engineering, Virginia Polytechnic Institute and State University, Blacksburg, VA 24060, USA

[4]Institute for Computational and Data Science, The Pennsylvania State University, University Park, PA, 16802

Corresponding authors: Saeed S.I. Almishal saeedsialmishal@gmail.com

## Abstract

We report a temperature-composition phase diagram for the chemically disordered and $CeO_2$ – $LA_2O_3$ high entropy oxides (HEOs), where LA denotes equimolar Y, La, Sm, and Pr, delineating stability regions for bixbyite, disordered fluorite, and intermediate vacancy-ordered fluorite phases. The diagram is constructed from a characterization package applied to bulk ceramics including X-ray diffraction (XRD), transmission electron microscopy (TEM) electron diffraction, Raman spectroscopy, energy-dispersive spectroscopy, X-ray absorption near-edge structure spectroscopy, and ultraviolet–visible spectroscopy, to quantify crystal structure at multiple length-scales, local coordination environments, and electronic structures across the formulation space. This comprehensive measurement suite is critical to identify boundaries between the closely related phases. For example, Raman scattering reveals local structural and defect environments unique to bixbyite local order that persist to ~50% Ce under equilibrium synthesis conditions but are invisible to XRD and TEM. We also report a companion thin film study to demonstrate that quenched kinetic energy from a physical deposition process can metastabilize the high symmetry, and thus high entropy, fluorite phase with only 20% Ce. This is noteworthy because electroneutrality constraints demand an exceptionally vacated oxygen sublattice; we estimate 16.7%, approaching that of δ-$Bi_2O_3$. Together, our equilibrium ceramics and far-from-equilibrium thin films show that when synthesis is coupled with rigorously chosen, multi-length-scale characterization, now one can identify the phase stability thermodynamic drivers and simultaneously derive practical guidelines for experimentally realizing targeted phases and structures – and thereby deliberately engineer properties in $CeO_2$-$LA_2O_3$ HEOs, whose broad defect chemistries demand such an approach.

# Introduction

Stabilizing $Mg_{1/5}Co_{1/5}Ni_{1/5}Cu_{1/5}Zn_{1/5}O$ into a single-phase rock salt above 850°C catalyzed high-entropy oxide (HEO) research in 2015.[1–5] This result positions configurational-entropy driven and configurational-entropy assisted routes as new strategies for accessing complex oxides with potentially emergent functions.[1,2,6,7] In 2017, Djenadic et al.[8] extended this approach to lanthanide sesquioxides ($A_2O_3$) and dioxides ($AO_2$), synthesizing the prototype lanthanide HEO, $Y_{1/5}La_{1/5}Ce_{1/5}Pr_{1/5}Sm_{1/5}O_{2-\delta}$ (hereafter equimolar F1), via nebulized spray pyrolysis (NSP)[8]. Their X-ray diffraction (XRD) results suggest that equimolar F1 crystallized into a single-phase defective fluorite structure (space group Fm-3m); they infer 16.5% oxygen vacant sites (i.e, $\delta \approx 0.33$), given that Y, La, and Sm remain in the 3+ oxidation state, Ce adopts a 4+ state, and Pr exhibits 2 $Pr^{4+}$:1$Pr^{3+}$ mixed valence.[8] Therefore, interest in this material class is predicted on the expectation that leveraging cation-valence control to engineer tunable and significant oxygen vacant sites promises pathways toward mass and charge transport applications spanning solid-oxide fuel cells to neuromorphic computing. However, a central challenge in designing and utilizing this material class is understanding and characterizing the structural nuances, phase evolution and kinetics that favor one structural phase or configuration over another. For example, equimolar F1 fluorite converts to the bixbyite-type $c$ phase ($Ia\bar{3}$) when sintered conventionally or when NSP-prepared fluorite is annealed in air. These transformations demonstrate that bixbyite is the equilibrium phase under these conditions, in accordance with zero-kelvin calculations that consistently identify cubic bixbyite as enthalpically favored over fluorite (See Supplementary Information Note 1). This cubic phase has an ordered oxygen sublattice and is consequently less appealing for mass-transport applications.[8] Other work shows that chemical doping can yield single-phase fluorite through solid-state synthesis by adding an equimolar sixth fluorite-stabilizing cation, such as $Zr^{4+}$ or $Hf^{4+}$, to create the cation-valence balance that stabilizes a defective-fluorite lattice, but at the expense of fewer oxygen vacant sites than equimolar F1.[9,10]

A unified framework linking thermodynamic stability, persistent metastable states, and synthesis kinetics, that respects the often-nuance differences separating distinct phases, is needed to contextualize high configurational entropy systems and to rationalize their observed and expected properties. So-doing, however, requires a significant characterization burden because the inherent chemical disorder often denies that one structural description applies across all relevant length scales. To address this gap in the $CeO_2$-$LA_2O_3$ system, we develop and report a collaborative study of bulk ceramics and thin films where Ce is systematically substituted with an equimolar lanthanide mixture of Y, La, Pr, and Sm; i.e., the prototypical five-component $Ce_x(YLaPrSm)_{1-x}O_{2-\delta}$ system where $x = 0.2$ for the highest configurational entropy and chemical disorder as illustrated in Figure 1(a). Importantly, this single compositional axis dimentionalizes the experiment for visualization as a pseudo-binary system. Ce concentration is a particularly powerful and clean control parameter to form this understanding because: (1) Ce-free (Y, La, Pr, Sm) compositions do not form single-phase cubic bixbyite or fluorite at laboratory temperatures, equimolar F1 with 20% Ce in forms global single-phase bixbyite, and $CeO_2$ itself stabilizes as fluorite; and (2) $Ce^{4+}$ is thermodynamically favored at ambient $pO_2$, and its oxygen chemical potential strongly overlaps with the lanthanides $3^+$ oxygen potentials, demanding that the many-cation fluorite lattice will host fully 8-coordinated $Ce^{4+}$ environments and on average ½ of the 3+ lanthanide cations in 7-coordination to maintain electroneutrality (assuming no electronic carriers).[11–14] In this limit, it is entropically favorable for the vacated oxygen positions to be randomly distributed, yielding what we refer to as a disordered defective-fluorite phase (or disordered vacated-oxygen-sublattice fluorite), in which both the cation sublattice and the vacated-sites-bearing oxygen sublattice are disordered (Details on how to handle this landscape computationally is available in Supplementary Information Note 1). Because these intertwined valence and structural complexities cannot be resolved through global-averaging techniques such as XRD alone, constructing a reliable equilibrium phase diagram requires rigorous, multi-length-scale characterization. This includes methods that directly probe local structure, chemical environments, vibrational modes, and electronic states.

At the same time, these complexities indicate that equilibrium processing captures only part of the accessible structural landscape in lanthanide HEOs, and the competing fluorite–bixbyite energetics suggest that non-equilibrium routes with kinetic stabilization may reveal metastable states that equilibrium methods suppress or cannot reach.[15] Pulsed laser deposition (PLD) provides a particularly powerful far-from-equilibrium route as it rapidly quenches a high-entropy plasma plume onto a relatively cold substrate, kinetically trapping high-energy cation configurations at temperatures far below those needed for bulk stabilization.[7,16,17] This kinetic trapping opens a complementary view of phase space – one shaped by composition and effective temperatures well beyond most material melting points. Together, these considerations underscore the need for a reliable equilibrium baseline that can only be known through rigorous multimode/multi-length-scale characterization: defining the lowest energy configurations is the critical first step to appreciate the stable and metastable macrostates that are accessible for property engineering.

## Results and discussion

### Constructing a comprehensive multiscale near-equilibrium thermodynamic phase diagram

We begin by constructing a comprehensive "global scale" equilibrium phase diagram for the $Ce_x(YLaPrSm)_{1-x}O_{2-\delta}$ system, varying Ce concentration from 20% to 80% across nine compositions. Using 24 sintering temperature experiments each repeated at least three times, we map Ce content on the x-axis and temperature on the y-axis, depicted in Figure 1(b). Samples were thoroughly milled, as detailed in Methods, to promote rapid and uniform phase formation and densification. To approximate near-equilibrium conditions, we employ a 10-hour sintering dwell for all experiments, a dwell duration which we validate by time-sensitivity tests showing that, after proper milling, the final phases can be reached in as little as 1 hour (Supplementary Information Figure S2). We identify phases by ex situ XRD and classify samples as fluorite (F-type) only when bixbyite (C-type) reflections are fully absent when viewed on a logarithmic scale (Figure 1(b-c)). In turn, the phase diagram in Figure 1(b) delineates three phase regions: multi-phase, single-phase bixbyite, and single-phase fluorite; and a fourth faded region that bridges the latter two.

We first examine compositions with varying Ce concentrations, all sintered at 1500 °C, their XRD patterns are shown in Figure 1(c). At 20% Ce, the material adopts a cubic bixbyite structure, as apparent by the full set of characteristic bixbyite peaks. The relative intensity of these reflections progressively decreases with increasing Ce content, and they vanish entirely above 35% Ce, marking a transition to a single-phase fluorite structure, likely with significant vacated oxygen sites to maintain global charge neutrality, i.e, single-phase disordered defective fluorite structure. Notably, density functional theory calculations at 0 K consistently predict bixbyite to be the enthalpically favored structure across all Ce concentrations (20–95%) and at different oxygen-vacancy levels (Supplementary Information Note 1). This reinforces that high temperatures, together with both cation and anion configurational entropy gain, are critical for stabilizing the high-entropy fluorite phase. These results identify collectively that the interval approximately between 32.5% and 35% Ce as the transition regime in which bixbyite-like and fluorite-like local environments likely coexist, under high temperature equilibrium synthesis, and some caution is needed during XRD analysis which only samples longer-range structure.

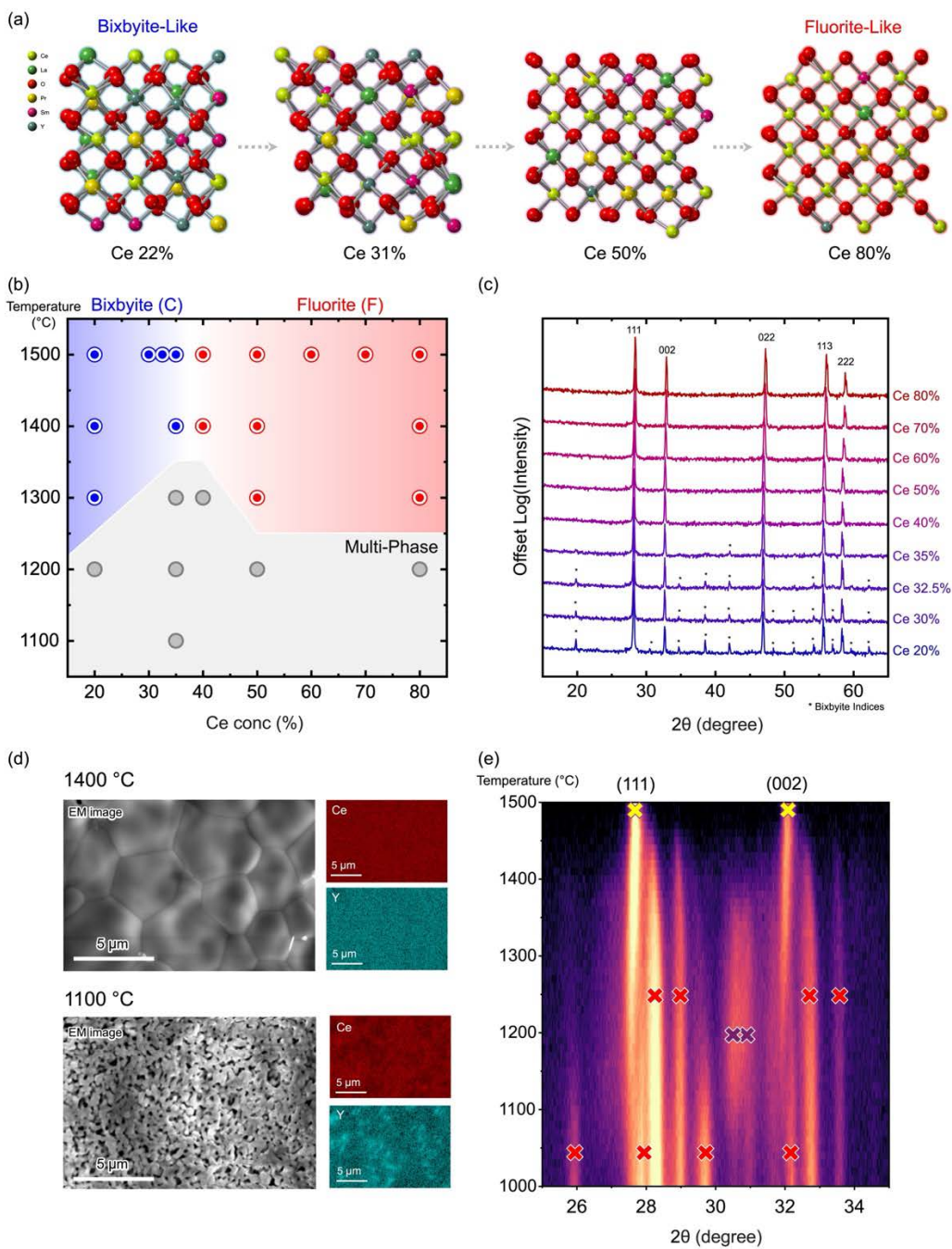

**Figure 1 | Macroscopic structural identification by X-ray diffraction. (a)** Illustration of the structural evolution of the for $Ce_x(YLaPrSm)_{1-x}O_{2-\delta}$ from the DFT simulated supercells (Supplementary Information Note 1). **(b)** Phase diagram for $Ce_x(YLaPrSm)_{1-x}O_{2-\delta}$, where blue circles, red circles, and grey circles denote the single bixbyite phase, single fluorite, and multi-phase, respectively. **(c)** X-ray diffraction patterns of $Ce_{1-x}(YLaPrSm)_xO_{2-\delta}$ (x=0.2~0.8) sintered at 1500 °C with 10-hour dwell time (* indicate the bixbyite characteristic peaks). **(d)** Scanning electron microscope images and elemental dispersive spectroscopy mapping of Ce and Y for the sintered $Ce_{0.325}(YLaPrSm)_{0.675}O_{2-\delta}$ pellets at 1100 and 1400°C with 10-hour dwell time. **(e)** In situ high temperature X-ray diffraction pattern of $Ce_{0.325}(YLaPrSm)_{0.675}O_2$ powder, , shown over a reduced 2θ range with indexed reflections; the full data set is provided in Supplemental Materials Note 5. Red crosses identify reflections from individual precursors, purple crosses demark intermediate phases, and yellow crosses demark the final fluorite phase characteristic peaks.

Exploring the individual transition temperatures in Figure 1(b) as a function of composition produces a self-consistent set of companion trends (Supplementary Information Note 3). At 1300 °C, the 20% Ce composition transitions from multiphase to single-phase bixbyite but never reaches fluorite under equilibrium synthesis, compositions with 32.5–40% Ce require an additional 100 °C to enter a single-phase thermodynamic boundary region with competing bixbyite and fluorite motifs, at 50% Ce composition single-phase defective fluorite becomes stable between 1200 °C and 1300 °C. SEM–EDS mapping (Supplementary Information Note 4) across all sintered compositions shows that all five cations remain uniformly distributed at the microscale, with no evidence of any additional elements beyond those in the F1 formulation. Furthermore, the Ce concentrations extracted from the EDS spectra closely match the values established during powder formulation (Supplementary Information Note 4 Table S1 and Figure S4(j)), demonstrating precise compositional control during synthesis. Figure 3(d) shows an important SEM and EDS comparison of ceramics sintered at 1100 °C and 1400 °C, the additional 300 °C is necessary for densification, cation homogenization, and grain growth, and for properly separating the kinetic and thermodynamic contributions to phase evolution.

To complement ex situ XRD measurements with additional kinetic information, we conduct an in situ high-temperature XRD study using the 32.5% Ce composition as a representative case. The entire data set is shown in the Supplementary Materials Note 5, a reduced two-theta range is shown with indices in Fig 1(e). Red crosses identify reflections from individual precursors, purple crosses demark intermediate phases, and yellow crosses demark the final fluorite phase characteristic peaks. Strongest reactions occur in the temperature region between 1100° C and 1300 °C, with complete or near-complete reaction by 1450 °C. This behavior differs slightly from the ex situ result for the 32.5% Ce composition, where weak bixbyite peaks remain visible even after sintering at 1500 °C. There are two possible, and probably co-contributing reasons: (i) ex situ scans have significantly higher signal-to-noise ration because they are longer and may better detect minor bixbyite remnants, and (ii) the ex situ samples undergo a 10-hour sintering dwell, followed by furnace cooling to 1000°C at 5 °C/min before direct quenching to ambient. It is possible that some fraction of material back-converts to bixbyite during this additional thermal history below the transition range.

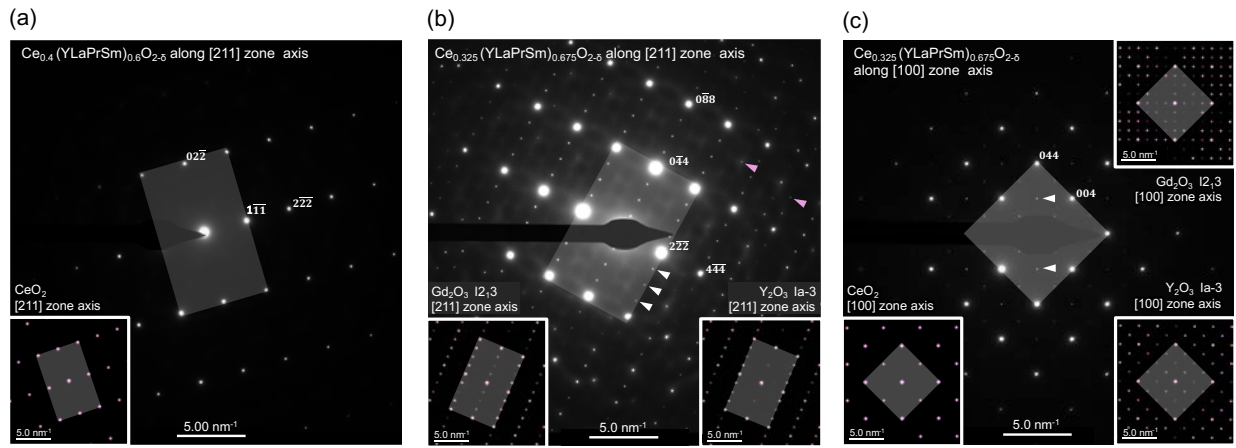

**Figure 2 | Local structural analysis by selected-area electron diffraction in the Ce-rich $Ce_x(YLaPrSm)_{1-x}O_{2-\delta}$ transition regime. (a)** SAED pattern for the 40% Ce sample, $Ce_{0.4}(YLaPrSm)_{0.6}O_{2-\delta}$, acquired along the [211] zone axis, with a simulated $CeO_2$ [211] pattern (inset), supporting assignment of fluorite $Fm\bar{3}m$ symmetry. **(b,c)** SAED patterns for $Ce_{0.325}(YLaPrSm)_{0.675}O_{2-\delta}$ along the **(b)** [211] and **(c)** [100] zone axes show satellite reflections relative to the ideal fluorite lattice. Insets show simulated patterns for $Gd_2O_3$ ($I2_13$) and $Y_2O_3$ ($Ia\bar{3}$) along the corresponding zone axes for comparison. White triangles in **(b)** mark satellites consistent with $Gd_2O_3$, while magenta triangles highlight additional ordering along ⟨111⟩ directions not captured by either reference; white triangles in **(c)** mark satellites consistent with $Y_2O_3$. The pronounced diffuse scattering in **(b)** is indicative of short-range ordering. Together, these data indicate that the 32.5% Ce composition lies within a structural transition regime where nanoscale symmetry-lowering ordering reminiscent of multiple lanthanide sesquioxides is superimposed on an average fluorite framework.

To further resolve the local structural environments in the transitional Ce-concentration regime, and to clarify the bixbyite space-group assignment, which remains debated in the literature[8,15], we perform TEM selected-area electron diffraction (SAED) on samples with 32.5% and 40% Ce. We previously reported the SAED pattern of the equimolar 20% Ce composition, which is consistent with a bixbyite-like structure closely approximating the $I2_13$ space group[15]. Figure 2(a) shows the SAED pattern for the 40% Ce sample along the [211] zone axis and a simulation of $CeO_2$ along the same direction. From this we can assign the $Fm\bar{3}m$ space group. Fig.s 2(b,c) show the SAED patterns for the 32.5% Ce sample along the [211] and [100] zone axes respectively, both show satellite spots, relative to the high symmetry fluorite structure, indicating additional periodicity and the [211] shows significant diffuse scattering that likely occurs due to short range order. Insets include simulations from $Gd_2O_3$ (monoclinic bixbyite space group $I2_13$) and $Y_2O_3$ (cubic bixbyite space group $Ia\bar{3}$) for comparison. White triangles in Fig. 2(b) identify satellite spots consistent with $I2_13$ $Gd_2O_3$, white triangles in 2(c) indicate satellite spots consistent with $Ia\bar{3}$ $Y_2O_3$, while magenta triangles in 2(b) indicate ordering from along [111] directions not captured in either. These results reinforce our interpretation that the 32.5% Ce composition exists within a structural transition zone with nanoscale regions featuring symmetry-lowering ordering reminiscent of multiple lanthanide sesquioxides and/or fluorites. It is important to note that the X-ray diffraction of these samples appear to have weak bixbyite reflections suggesting that the ordered regions are either too small, or they involve local ordering on the oxygen sublattice within a fluorite lattice, in either case they would be challenging to detect.

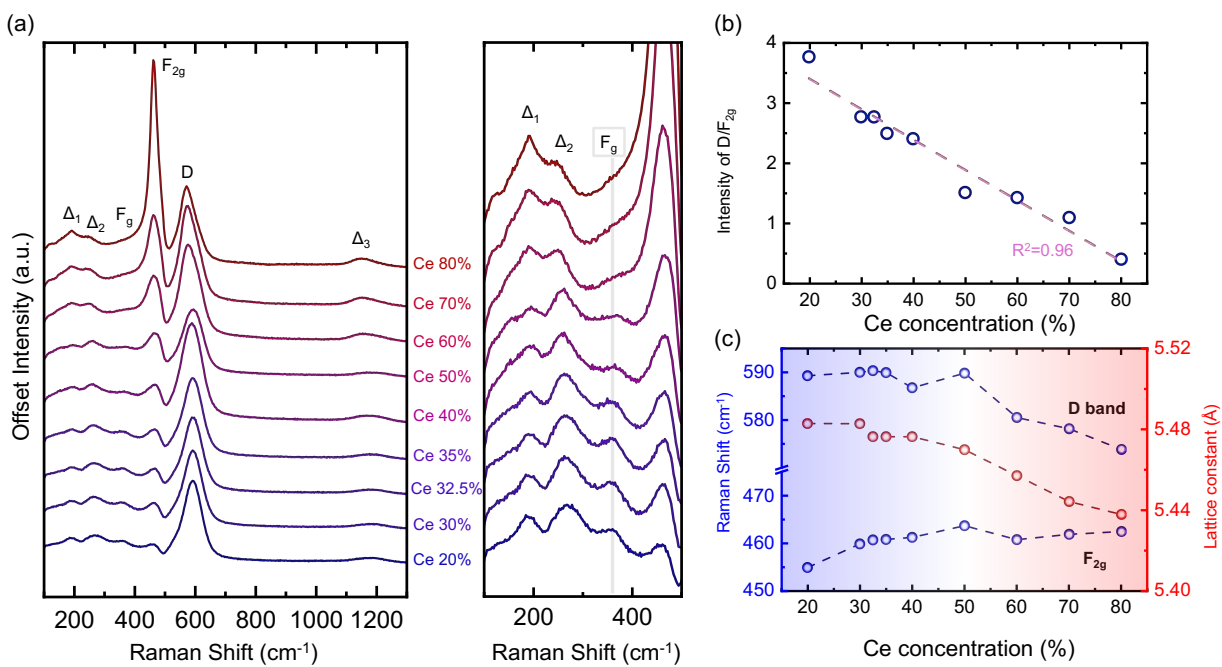

**Figure 3 | Structural and defect analysis of the $Ce_x(YLaPrSm)_{1-x}O_{2-\delta}$ system by Raman spectroscopy.**
**(a)** Raman spectra of $Ce_x(YLaPrSm)_{1-x}O_{2-\delta}$ pellets ($x$ = 0.2–0.8). The left panel plots the Raman shift from 100 to 1300 cm$^{-1}$, while the right panel provides a zoomed view of the 100–500 cm$^{-1}$ region. **(b)** Intensity ratio of the defect-related D band to the fluorite $F_{2g}$ mode (D/$F_{2g}$) as a function of Ce concentration. **(c)** Raman shifts of the D band and $F_{2g}$ mode plotted against the corresponding lattice parameters (bixbyite lattice constants divided by 2 for direct comparison with fluorite-type samples).

To more sensitively probe the local structural environments and complement the XRD and TEM analyses, we employ Raman spectroscopy, which is highly sensitive to oxygen–cation vibrational modes and particularly effective at detecting subtle changes at short length scales in the doped ceria.[18–20] Figure 3a shows the Raman spectra of nine ceramic samples with the same compositions as in the phase diagram in Figure 1(b), all sintered at 1500 °C, the left panel plots the Raman shift on the x-axis from 100 to 1300 cm$^{-1}$, and the right panel provides a zoomed view from 100 to 500 cm$^{-1}$ (All Raman shifts are summarized in Supplementary Information Note 6). The $\Delta_1$, $\Delta_2$, D, and $\Delta_3$ bands at ~190, 260, 560, and 1170 cm$^{-1}$ represent overtone or secondary-scattering features[21]. The $\Delta_1$, $\Delta_2$, D, and $\Delta_3$ bands at ~190, 260, 560, and 1170 cm$^{-1}$ represent overtone or secondary-scattering features[21]. The broader $\Delta_1$–$\Delta_2$ and D bands likely originate from local symmetry breaking, structural distortions, or secondary phases as Ce concentration changes.[21] The broad $\Delta_3$ band features around 1180 cm$^{-1}$ represents the electronic transition excited by the Raman laser in ceria systems[19]. The primary peaks, however, near 460 cm$^{-1}$ reflects the $F_{2g}$ mode of the 8-coordinated fluorite structure, while the 355 cm$^{-1}$ peak reflects the $F_g$ mode of the 6-coordinated bixbyite structure[22]. A prominent $F_g$ signature is present in the 20% Ce sample, consistent with the cubic bixbyite phase identified by XRD and TEM. Its intensity decreases with increasing Ce concentration and vanishes above 50%, which is higher than the ~35% and ~40% Ce thresholds inferred from XRD and TEM respectively. Further insight comes from the D band, which reflects structural defects such as oxygen vacancies and symmetry-lowering distortions, and its intensity relative to the $F_{2g}$ peak provides a quantitative disorder metric, which we plot in Figure 3(b). The D band is strongest at low Ce concentrations and weakens steadily with increasing Ce, whereas the $F_{2g}$ peak strengthens, indicating that Ce addition progressively reduces structural defects – here corresponding to vacated oxygen lattice sites. In addition to the intensity trends, we observe systematic peak shifts with increasing Ce content. The $F_{2g}$ mode shifts to higher wavenumbers, whereas the D band shifts to lower wavenumbers; both trends correlate with the decrease in lattice parameter measured by XRD, as shown in Figure 3(c). We attribute the $F_{2g}$ mode blue shift to a likely reduction in Ce–(Y,La,Pr,Sm) bond length and the corresponding lattice contraction, consistent with arguments made for simpler systems in Ref.s (23) and (24). Similarly, the D band red shift reflects changes in the vacancy-cluster dynamics: as Ce reduces the number of vacated oxygen lattice sites clusters, these clusters contract the near-neighbor coordination shell, driving the defect-related modes to lower wavenumbers.

This Raman analysis indicates that the family of similar, yet distinct structures experimentally resolved by diffraction likely emerges from a continuum of defect configurations, with cations becoming increasingly under- or over-coordinated relative to the fluorite and bixbyite endmembers, respectively. The chemical disorder in high-configurational-entropy materials creates a forgiving structural energy landscape, with phase boundaries smeared across composition space. Importantly, this structural study exemplifies the critical role of multimode characterization in understanding high-configurational-entropy systems. Information from local point defects and from short- and long-range order must be integrated to develop an accurate structural representation and physical understanding.

Correlating Phase Stability to the Local Electronic Structure

To connect our defect-based structural model to electronic configuration we next probe the electronic structure across the $Ce_{1-x}(YLaPrSm)_xO_{2-\delta}$ series using UV–vis diffuse reflectance spectroscopy, X-ray absorption spectroscopy, and a targeted reversibility study.

We start with probing the electronic structure using UV–vis diffuse reflectance. Tauc analysis[25] (Details in Supplementary Information Note 7) shows that all formulations exhibit a ~2 eV direct bandgap, this ~ 1 eV reduction compared to $CeO_2$ is attributed to $Pr^{4+}$.[26] The uniformity of this bandgap across the series suggests only a minimal influence from structure evolution. This motivates a direct interrogation of oxidation states and local electronic environments using X-ray absorption. We therefore examine the lanthanide $L_3$ edges and the Y K edge using X-ray absorption near edge structure spectroscopy (XANES) – all the results are compiled in (Supplementary Information Note 8 and Table S2)[13]. In Figure 4(a) we show only the Ce spectra to confirm that that Ce remains predominantly in the +4-oxidation state across the compositional series. In Supplementary Information Figure S8, we show that Sm, La, and Y remain trivalent, matching their starting parent oxide precursors[13]. Pr, however, exhibits mixed +4/+3 valence which aligns with the reduced optical bandgap compared to $CeO_2$.[13,26] From a defect-chemistry perspective, a persistent $Ce^{4+}$ state across the series implies a progressive reduction in vacated oxygen lattice sites concentration as Ce increases. Therefore, assuming $Ce^{4+}$, $Pr^{3.67+}$, $Y^{3+}$, $La^{3+}$ and $Sm^{3+}$, we plot in Figure 4(b) the expected vacated oxygen sites concentration with increasing Ce content. This trend complements the Raman results: the defect-related D-band intensity decreases steadily with higher Ce content, demonstrating that Ce addition simultaneously stabilizes the defective disordered fluorite phase and suppresses the corresponding vacated oxygen lattice sites-rich environments. Together, these findings shape the broader thermodynamic landscape of this lanthanide HEO system.

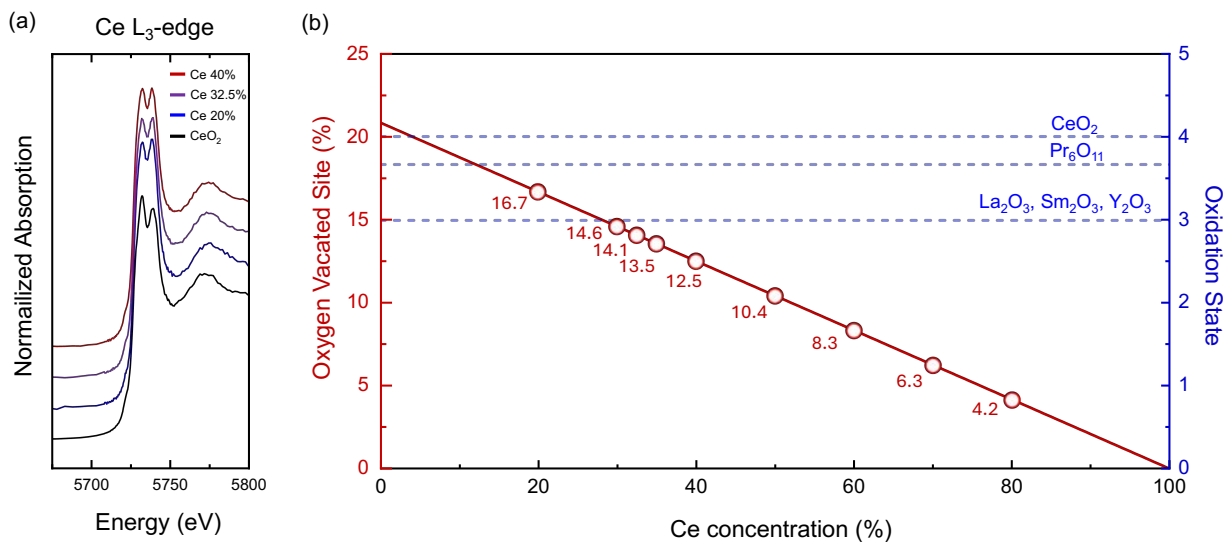

**Figure 4 | Electronic structure analysis and charge-balance–driven evolution of vacated oxygen sites of the $Ce_x(YLaPrSm)_{1-x}O_{2-\delta}$ system.** (a) Ce $L_3$-edge XANES spectra for $Ce_x(YLaPrSm)_{1-x}O_{2-\delta}$ powders with x = 0.20, 0.325, and 0.40 compared to a $CeO_2$ standard, showing that Ce remains predominantly in the 4+ oxidation state across the series. (b) Expected fraction of vacated oxygen lattice sites as a function of Ce content, calculated assuming $Ce^{4+}$, $Pr^{3.67+}$, and trivalent Y, La, and Sm. Open circles corresponds to compositions from the phase diagram in Figure 1(b), while blue dashed lines indicate the formal oxidation states of $CeO_2$, $PrO_2$, and the trivalent lanthanide oxides for reference. Increasing Ce content progressively reduces the required population of vacated oxygen lattice sites, consistent with the suppression of defect-rich environments inferred from Raman spectroscopy. Part (a) is regrouped and reproduced from our XANES focused manuscript: G. R. Bejger, M. K. Caucci, S. S. I. Almishal, B. Yang, J. Maria, S. B. Sinnott and C. M. Rost, *J. Mater. Chem. A*, 2025, 13, 29060 DOI: 10.1039/D5TA03815D.

If this so far reasoning holds, and in direct analogy to valence-stabilization behavior recently reported for rock salt oxides[12], a global fluorite composition such as the 40% Ce sample should switch its structural motif under redox cycling. Specifically, reducing the material should drive some Ce into the $Ce^{3+}$ state, which favors the ordered oxygen sublattice of the bixbyite phase; re-oxidizing the same sample should drive $Ce^{3+}$ back to $Ce^{4+}$ and restore the disordered oxygen-vacancy framework that stabilizes the global fluorite structure, as graphically depicted in Figure 5(a). We illustrate this reversible valence–structure coupling by comparing x-ray diffraction and Raman scattering spectra for bulk samples cycled between oxidizing and reducing conditions[27]. The reduced F40 (40% Ce) sample undergoes a clear transition from fluorite to bixbyite, evidenced by the emergence of bixbyite reflections in XRD (Figure 5b) and the strengthened $F_g$ mode in the Raman spectra (Figure 5c). Consistent with these structural changes, X-ray photoelectron spectroscopy measurements Figure 5(d) show a significant $Ce^{3+}$ population after forming-gas treatment. Upon re-oxidation XPS, XRD, and Raman features for $Ce^{3+}$, bixbyite ordering, and $F_g$ defect modes respectively are eliminated (The corresponding SEM EDS and the full XPS spectra are in Supplementary Information Note 9).

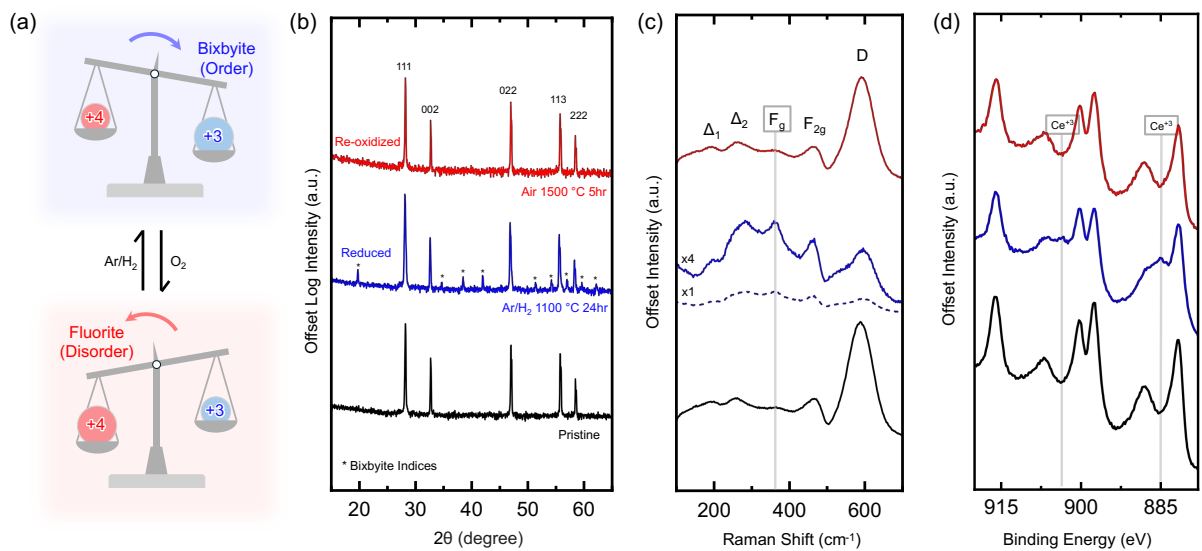

**Figure 5 | Phase reversibility of $Ce_{0.4}(YLaPrSm)_{0.6}O_{2-\delta}$. (a)** Schematic illustration of the bixbyite–fluorite phase transition mechanism in the Ce-driven HEO. **(b)** X-ray diffraction patterns of $Ce_{0.4}(YLaPrSm)_{0.6}O_{2-\delta}$ annealed under different gas environments (* symbols denote bixbyite characteristic peaks). **(c)** Raman spectra highlighting the corresponding structural evolution (x4 indicate quadruple the intensity of x1). **(d)** X-ray photoelectron spectra of the Ce 3d region, showing changes in Ce oxidation states following annealing under different gas conditions.

These results reinforce a model where $Ce^{4+}$ preferentially stabilizes the fluorite framework in bulk samples when concentrations are 35% or higher. To understand how this balance shifts during far from equilibrium synthesis, we next explore pulsed laser deposition (PLD) where the adatom kinetic energy can contribute to an effective $TS$ term and expand the accessible configurational space.

## Kinetically-arresting far-from-equilibrium microstates with different oxygen vacant sites ordering

We expect the defective fluorite structure, with its high configurational disorder on both cation and anion sublattices, to be entropically favored at high enough effective temperatures[15,28]. PLD plasmas contain significant adatom fractions with effective temperatures between 10,000–100,000 K, this combined with rapid quenching onto a comparatively cold substrate enables the access and the kinetic arrest of highly-metastable microstates that equilibrium routes suppress[17,29,30]. Consequently, we hypothesize that even compositions such as the equimolar F1 sample with only 20% Ce, which adopt bixbyite under equilibrium processing, can be realized metastably by physical vapor deposition methods like PLD. Importantly, PLD is compatible with a large total pressure range thus the capability to regulate deposition energetics by gas phase scattering combined with variable laser fluence and working distance[16,17]. Here, we demonstrate this using two growth parameters: laser fluence, which controls the synthesis energetics, and substrate temperature, which governs the quenching kinetics, to generate a far-from-equilibrium phase diagram mapped in Figure 6(a).

Figure 6 maps the XRD-identified phases for 120–140 nm F1 thin films as a function of laser fluence (x-axis) and substrate temperature (y-axis). We deposit films either directly on YSZ (001) or on a 15 nm $CeO_2$ buffer layer that reduces the ~7% lattice mismatch between the F1 film and the substrate. The full set of XRD θ–2θ patterns are summarized in Supplementary Information Figure S11, while Figure 6(b) highlights three representative films. We track bixbyite-like ordering with the (006) reflection, which originates from the doubled bixbyite lattice parameter and is absent in fluorite. Films grown at low laser fluence and high substrate temperature show a strong (006) peak, as in representative film 1 grown at 1.5 $J/cm^2$ and 650°C with the $CeO_2$ buffer. In contrast, films grown at high fluence and low temperature suppress this peak, consistent with the global fluorite-like structure observed in representative film 2 grown at 2.9 $J/cm^2$ and 450°C with the buffer. Although high-energy, low-temperature growth without the buffer yields the same conclusions with films that appear as single-phase defective fluorite, we often observe a minor (111) orientation, which we attribute to epitaxial strain arising from the substrate–film mismatch. Notably, all films grow smooth with step and terrace morphology as shown in the atomic force microscopy (AFM) micrographs in Figure S12.

The low signal from thin films makes Raman analysis interpretation difficult due to the limited scattering volume. Therefore, in thin films electron diffraction offers the best observation into structure and anion ordering (complementary EDS for the three films in Figure 6(b) is in Supplementary Information Note 10, Figures S13-15). Figures 6(c) shows the SAED pattern along the [110] zone axis for the low-fluence, high-substrate-temperature film (film 1) that appears to be bixbyite from XRD with an inset of simulated pattern for $I2_13$ $Gd_2O_3$ bixbyite model. While the relative intensity match is imperfect, ordering along [001] and [110] directions are captured, supporting the conclusions from XRD. Figure 6(d) shows the same [110] zone axis SAED for the high-fluence, low-temperature film (film 2) that appears to be fluorite from "global scale" XRD with an inset of simulated $CeO_2$ $Fm\bar{3}m$ space group. The experimental TEM pattern, however, contains weak satellite spots indicating partial ordering along [110], the same trend is also observed for sample 3 without a buffer layer in Supplementary Information Figure S15. Following the analysis of Han et al.[31], we attribute this pattern to a defective disordered fluorite phase with vacated oxygen sites with an energetic preference to populate {110} planes. Notably, this does not preclude the presence of local bixbyite-like ordering within the broader disordered anion sublattice framework.[32] Overall, these results show that PLD-grown films retain some degree of anion ordering, yet the disorder–order balance remains tunable through growth energetics and substrate temperature.

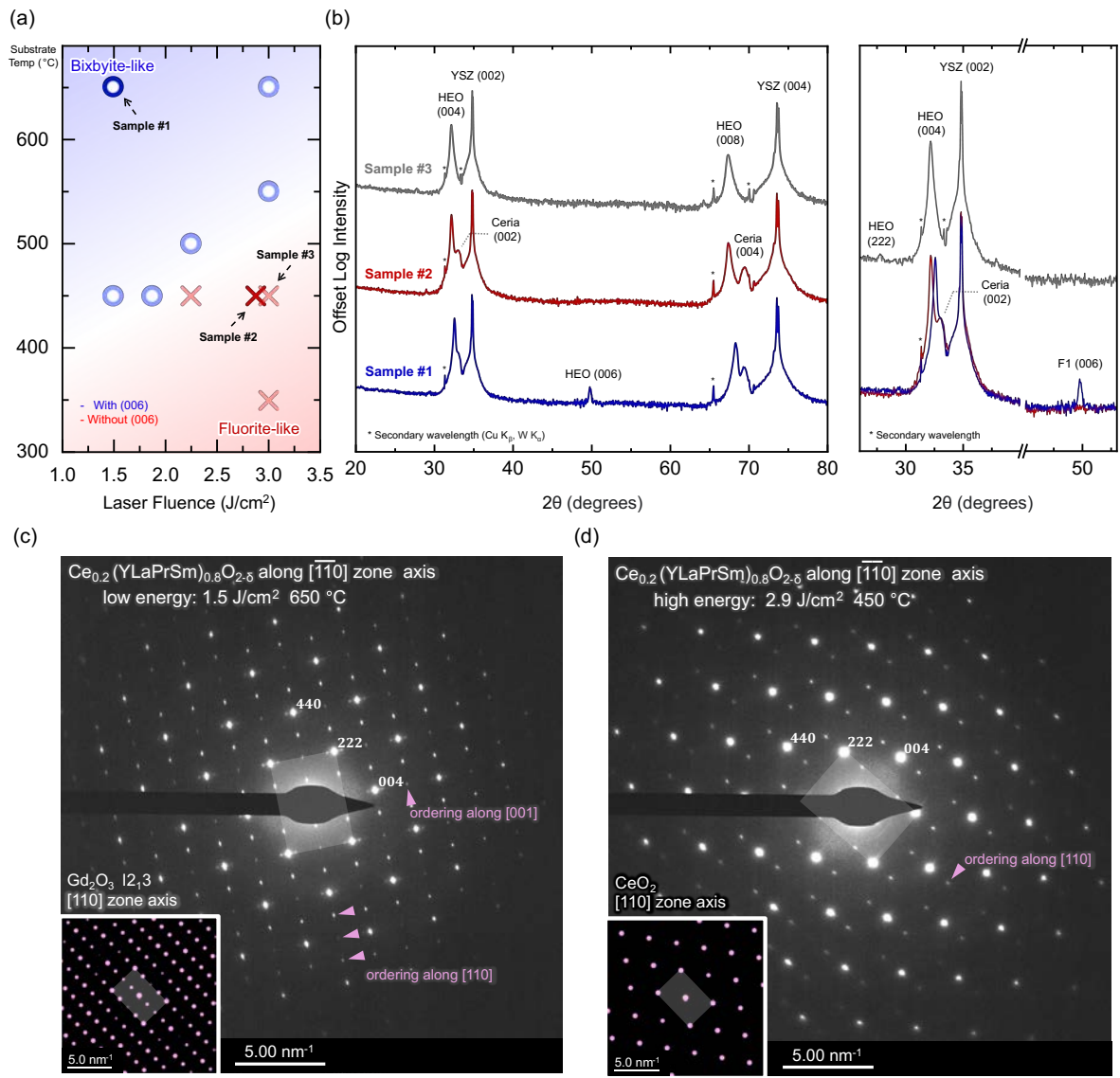

**Figure 6 | Thin-film growth of $Ce_{0.2}(YLaPrSm)_{0.8}O_{2-\delta}$ by pulsed laser deposition. (a)** Growth map for $Ce_{0.2}(YLaPrSm)_{0.8}O_{2-\delta}$, where blue circles and red crosses indicate X-ray diffraction patterns with and without the bixbyite 006 signature peak, respectively. **(b)** X-ray diffraction patterns of three representative films. Sample #1 (film 1), grown at 1.5 J cm$^{-2}$ and 650 °C on a CeO$_2$ buffer, exhibits a clear 006 peak. In contrast, sample #2 (film 2), grown at 2.9 J cm$^{-2}$ and 450 °C with the same buffer, suppresses this peak, consistent with its globally fluorite-like structure. Sample #3 is similar to sample 2 but grown without a buffer shown for additional comparison. **(c)** SAED pattern along the [110] zone axis for sample #1, with inset showing a simulated $I2_13$ Gd$_2$O$_3$ bixbyite pattern. Magenta triangles along [001] and [110] directions confirm bixbyite-like ordering consistent with the XRD results. **(d)** [110] SAED pattern for sample #2, with inset showing a simulated CeO$_2$ $Fm\bar{3}m$ fluorite pattern. Here, only weak satellites along ⟨110⟩ are observed, indicative of a defective disordered fluorite phase. Together, these data show that the balance between anion disorder and ordering in F1 films can be tuned by the PLD growth energetics and substrate temperature.

## A comment on functional properties: Ionic conductivity

Together, these findings define the structural–thermodynamic–kinetic landscape that governs fluorite stability across Ce-rich Lanthanide-HEOs. To connect this landscape to functional properties, we evaluate oxide-ion transport. Prior reports show that Ce-, Zr-, and Hf-stabilized high-entropy fluorites typically exhibit modest ionic conductivities of $\sim 3\times 10^{-4}$ S cm$^{-1}$ at 600 °C, well below benchmark electrolytes such as YSZ and GDC.[9,33] In contrast, electrochemical impedance spectroscopy on our bulk 20%, 32.5%, 50%, and 80% Ce compositions yields substantially higher conductivities of $3\times 10^{-3}$-$6\times 10^{-3}$ S cm$^{-1}$ at 600 °C, comparable to 10% GDC and to YSZ and an order of magnitude larger than any other reported Lanthanide-HEOs (Supplementary Information Note 11).[33,34] Activation energies extracted from Arrhenius fits follow the structural trends: bixbyite-rich compositions exhibit lower barriers (~0.6 eV) than fluorite-rich ones (~0.8 eV), a difference we attribute to enhanced electronic contributions from Pr in the bixbyite-like compositions.[35] These results reinforce that Ce content and oxidation state govern not only the structural motif and corresponding defect environments but also the ion-transport landscape. With targeted compositional design and further investigation into defect chemistry, oxygen-vacancy ordering, and local structural response, Lanthanide-HEOs present a promising and tunable platform for optimizing ionic conductivity.

In summary, our findings reveal that increasing Ce$^{4+}$ concentration lowers the fluorite formation enthalpy and enables the fluorite phase to stabilize through both cation and anion sublattice entropy. Even at low Ce$^{4+}$ levels, applying high kinetic energy with high effective temperature through PLD allows us to access this high configurational entropy landscape and kinetically arrest a global defective fluorite phase. Importantly, we have shown that no single characterization technique suffices on its own: the synergy of XRD, SEM, EDS, TEM, Raman, XANES, UV-Vis and XPS is essential to fully map the structural and defect landscape. This comprehensive approach is the key to navigating this complex structural landscape and potentially optimizing these materials for ionic conductivity, memristors, and neuromorphic computing applications.

## Methods

### Bulk Ceramic Synthesis

Bulk high-entropy oxide samples were synthesized using conventional powder processing techniques. The precursor powders $Pr_6O_{11}$ (Sigma-Aldrich 99.9%), $La_2O_3$ (Alfa Aesar 99.99%), $Sm_2O_3$ (Kurt Lesker 99.9%) were preheated at 700 °C for 12 hours to realize a single-phase starting precursors. The desired formulations were then weighed and mixed from these precursors in a addition to $CeO_2$ (Sigma-Aldrich 99.9%), and $Y_2O_3$ (Alfa Aesar 99.9%) powders. The mixed powders were then milled with 2 mm, 3 mm, and 5 mm yttrium-stabilized zirconia spherical media in a methanol slurry for 48 hours. The powder was then dried with a rotary evaporator and sieved (170-micron sieve). The sieved powders were then pressed into pellets 1.27 cm or 2.5 cm in diameter with 7500 lbs. using a Carver Laboratory Press, holding at the maximum pressure for 1 minute. Following this, the pellets were solid-state reacted in a box furnace to form a reactively-sintered bulk ceramics. The heating rate in the box furnace was set to 10 °C/min until 1000 °C and 5 °C/min thereafter to the targeted temperature. After the dwell time, the samples were then cooled at 5 °C/min until 1000 °C then quenched to room temperature.

### Bulk Ceramic Characterization

X-ray diffraction (XRD) was employed to characterize the crystal structure of the samples using a Panalytical Empyrean diffractometer equipped with a Cu X-ray source and Bragg-Brentano HD incident optics. In situ high-temperature XRD was performed at a heating rate of 10 °C/min using a Pt strip (HTK 2000N). Microstructure images and elemental distribution maps using energy-dispersive spectroscopy (EDS) were taken on a scanning electron microscope (SEM) Verios G4 with conducting carbon layers deposited on the surface. Raman data were collected in SWIFT mapping mode with a total mapping area of 20x20 μm$^2$ and step size of 1 μm, using a 633 nm laser and a 300 grooves/mm$^{-1}$ grating on the Horiba LabRam Raman system. The Raman shift position were fitted with Lorentz and Gaussian functions for the F2g peak and D band, respectively.[22] UV-Vis data was obtained by an Agilent Cary 5000 with a Praying Mantis diffuse reflectance accessory. The sintered pellets (dark brown color) are ground and diluted with KBr to enhance reflectance. The optical bandgap is determined by using the Tauc relation and the Kubelka–Munk function[25]. X-ray photoelectron spectroscopy (XPS) spectra were obtained using a monochromatic Al Kα X-ray source (hv = 1486.7 eV) with a pass energy of 27 eV for Y and 55 eV for the rest of the elements on Physical Electronic Versa Probe II. Energy calibration was done using the C 1s signal. X-ray absorption near-edge structure (XANES) results were measured at beamline 12-BM-B of the Advanced Photon Source, Argonne National Laboratory. The samples were measured in fluorescence mode with a Hitachi Vortex-ME7 silicon drift detector and processed using the Demeter package for XAS analysis[36]. Complete XANES analysis is available in our XANES focused manuscript in Ref ([13]).

Selected area electron diffraction (SAED) using transmission electron microscopy (TEM) was performed to investigate the local crystal structure of the samples at varying Ce concentration (32.5% and 40%). The experiments were conducted using Thermo Fisher Talos F200X at 200 kV accelerating voltage. Sample preparation for the TEM experiments was carried out using Thermo Fisher Helios focused ion beam (FIB), and a lamella was extracted parallel to the cleaved edge of the pellet.

The reducing environment experiment in the reversibility study was carried in a 5 cm diameter quartz tube furnace (Across International NC2156188). We seal the furnace and flow 100 SCCM of Linde UHP Ar + 1%$H_2$, controlled by a Brooks mass flow controller, at 1100 °C for 24 h. After the hold, we begin cooling still under continuous Ar flow, decreasing the temperature at 40 °C/min down to 700 °C, then quenching to air.

Electrochemical impedance spectroscopy (EIS) measurements were performed with a Modulab XM potentiostat on the bulk samples with 8 mm diameter sputtered Pt circle electrodes. Impedance was measured across frequencies ranging from 0.1 Hz to 100 kHz in the intermediate temperature range (400–700°C).

Thin Film Deposition and Characterization

Thin films were deposited by pulsed laser deposition (PLD) with a KrF excimer laser (248 nm, Coherent) using the ceramic targets made from bulk ceramic processing. YSZ (Shinkosha) substrates were sonicated in IPA for 15 min (step YSZ (001) substrate was obtained by annealing at 1200 °C in air for 10 hr). The substrates were then transferred to the PLD chamber at 200°C heater temperature and subsequently heated to the desired substrate growth temperature (350-650 °C). During growth, 50 sccm of $O_2$ was flown into the chamber, and the chamber pressure was adjusted to 50 mTorr total pressure. Laser fluence was varied between 1.5-3 J/cm$^2$, and the total deposition time was fixed at 200 sec for high-entropy oxide growth at 5-10 Hz. A ceria buffer layer (~15 nm) is grown at a substrate temperature of 650 °C using a laser fluence of 1.5 J/cm$^2$. After growth, all films are cooled to 150 °C in the chamber before transferring to the load lock and then to air. XRD was performed with a Cu X-ray source and Bragg-Brentano HD incident optics, and the surface morphology was characterized using an Asylum Research atomic force microscope (AFM). SAED patterns for the thin-film samples were acquired under conditions similar to those used for the bulk samples, using the same Thermo Fisher Talos X200 microscope. Sample preparation was carried out following the same procedure as for the bulk samples. The TEM lamella was extracted at a 45° angle relative to the substrate edge.

## Acknowledgements

The authors gratefully acknowledge the full support from NSF MRSEC DMR-2011839. The authors also acknowledge George N. Kotsonis for his assistance in the early development of this project, and Wes Auker, Nichole Wonderling, and the Penn State Materials Characterization Lab for their support with TEM sample preparation and high-temperature XRD measurements, respectively. The authors further acknowledge many insightful discussions with the members of the Penn State NSF MRSEC IRG2: Crystalline Oxides with High Entropy.

## Supplementary Information:

### Note 1: Density Functional Theory Calculations

To evaluate the role of Ce in stabilizing either the fluorite or bixbyite structure, we performed DFT calculations for compositions over a range of Ce concentrations (x) and oxygen non-stoichiometries ($\delta$), and constructed convex hulls of formation enthalpy $\Delta H_f$ for both phases.[37] **Figure S.1** summarizes the minimum formation enthalpies for fluorite (random $V_O$ configurations) and bixbyite (ordered $V_O$ configurations). For each Ce concentration, only the lowest-energy $\Delta H_f$ among the considered $V_O$ concentrations is plotted; these minima form the solid red (fluorite) and dashed blue (bixbyite) convex-hull lines. Because the optimal $\delta$ differs between both phases, we additionally include cross-evaluated points to enable comparison at fixed vacancy content. The red diamonds with dashed red hull represent fluorite $\Delta H_f$ values evaluated at the $\delta$ value corresponding to the lowest-energy bixbyite composition at that Ce concentration. Similarly, the blue circles with blue solid hull show bixbyite $\Delta H_f$ values evaluated at the $\delta$ value corresponding to the lowest-energy fluorite composition.

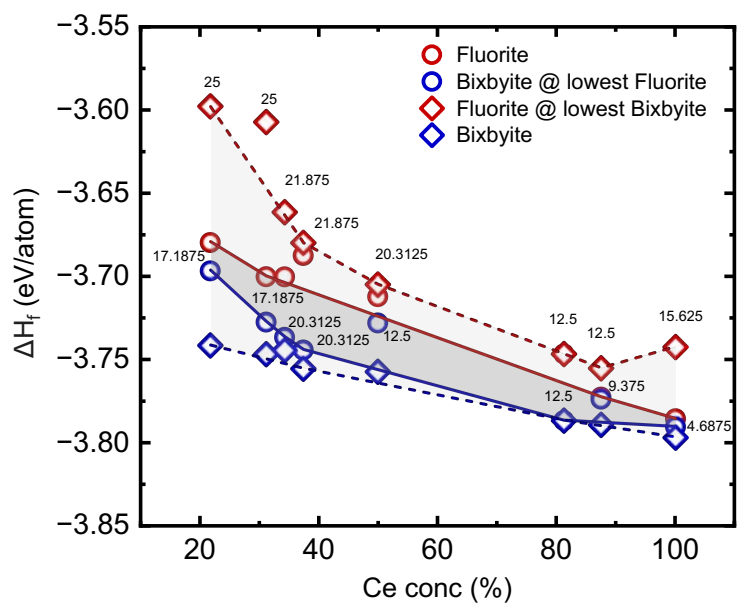

**Figure S1**: DFT calculated minimum formation enthalpies ($\Delta H_f$) of $Ce_x(YLaPrSm)_{1-x}O_{2-\delta}$ as a function of Ce concentration. Red circles and solid line show the minimum $\Delta H_f$ for fluorite structures (random $V_O$), and blue diamonds and dashed line show the minimum $\Delta H_f$ for bixbyite structures (ordered $V_O$). For each Ce concentration, only the lowest-energy $\Delta H_f$ among the considered $\delta$ values is plotted. Cross-evaluated points are included for comparison at matched vacancy contents: red diamonds (red dashed line) represent fluorite $\Delta H_f$ evaluated at the $\delta$ corresponding to the lowest-energy bixbyite composition, while blue circles (blue solid line) represent bixbyite $\Delta H_f$ evaluated at the $\delta$ corresponding to the lowest-energy fluorite composition. Data points label $\delta$ values.

**Computational Methods**

First-principles calculations were performed using the projector augmented-wave (PAW) method as implemented in the 6.3.0 version of the Vienna ab initio simulation package (VASP) software[38,39]. For the exchange correlation functional, we employed the r²SCAN meta-GGA functional[40], as it has showed strong performance for rare-earth oxide systems[41]. For the plane-wave basis, the kinetic energy cutoff was set to 700 eV, and the electronic self-consistency convergence threshold was achieved with an energy criterion of $10^{-6}$ eV. Ionic relaxations iterated until all atomic forces were less than 0.02 eV/Å. Structural optimization used the conjugate gradient algorithm from setting ALGO to All. Γ-centered k-point meshes of 2×2×2 were generated automatically using a KSPACING of 0.4 Å$^{-1}$. The pseudopotentials Ce, La, Pr, Sm, Y_sv, and O were selected from the PAW 64 dataset, with the *f* electrons of Ce, Pr, and Sm explicitly treated as valence electrons. All calculations assumed initial ferromagnetic spin alignment.

To model the random oxide compositions, seven 96-atom fluorite supercells ($Fm\bar{3}m$) were constructed using 2×2×2 conventional unit cells of Ce$_x$(YLaPrSm)$_{1-x}$O$_2$ with Ce concentrations x = 0.22, 0.31, 0.34, 0.38, 0.50, 0.81, and 0.88. The 32-site cation sublattices in each supercell were populated using the special quasi-random structure (SQS) approach[42], implemented via the Integrated Cluster Expansion Toolkit (ICET)[43]. For each Ce concentration, identical cation configurations were applied to generate the corresponding bixbyite ($Ia\bar{3}$) supercells, which also contain 32 cations per conventional unit cell. For each composition, several oxygen vacancy concentrations V$_O$ (non-stoichiometries δ) were examined. In fluorite structures, oxygen vacancies were introduced randomly, whereas in bixbyite structures the oxygen vacancies were restricted to the Wyckoff *8b* positions, in accordance with the crystallographic site symmetry. Except for 0% V$_O$ and 25% V$_O$, each composition has at least two anion configurations to account for effects of local vacancy arrangement.

The enthalpy of formation ΔH$_f$ was calculated from the DFT total energies. The equation for ΔH$_f$ is taken as the energy released when a RE-HEO compound is formed from the elemental constituents in their standard states[44]:

$$\Delta H_f^{298K} \approx \Delta H_f^{0K} = E_{tot} - \sum_{x_i} x_i \mu_i$$

Where $\Delta H_f^{0K}$ is the DFT calculated formation enthalpy, $E_{tot}$ is the total DFT energy of the RE-HEO containing $x_i$ atoms of element *i*, which has an elemental chemical potential of $\mu_i$ per atom.

## Notes 2: Dwell time stability experiments by both in situ and ex situ X-ray diffraction

Figure S2 (a,b) show the in situ high temperature X-ray diffraction of a pre-sintered (1500 °C, 1 hr) $Ce_{0.325}(YLaPrSm)_{0.675}O_2$ pellet annealed for an addition 1.5 hr in situ at 1500 °C. Figure S2 (c) shows the ex situ X-ray diffraction pattern of $Ce_{0.325}(YLaPrSm)_{0.675}O_2$ pellets sintered with different dwelling times from 1 hr to 24 hr at 1500 °C. Both in situ and ex situ XRD patterns show little to no time dependence at our measured durations beyond 1 hour of sintering when it comes to global crystal structure and confirm the phase stability of the synthesized material.

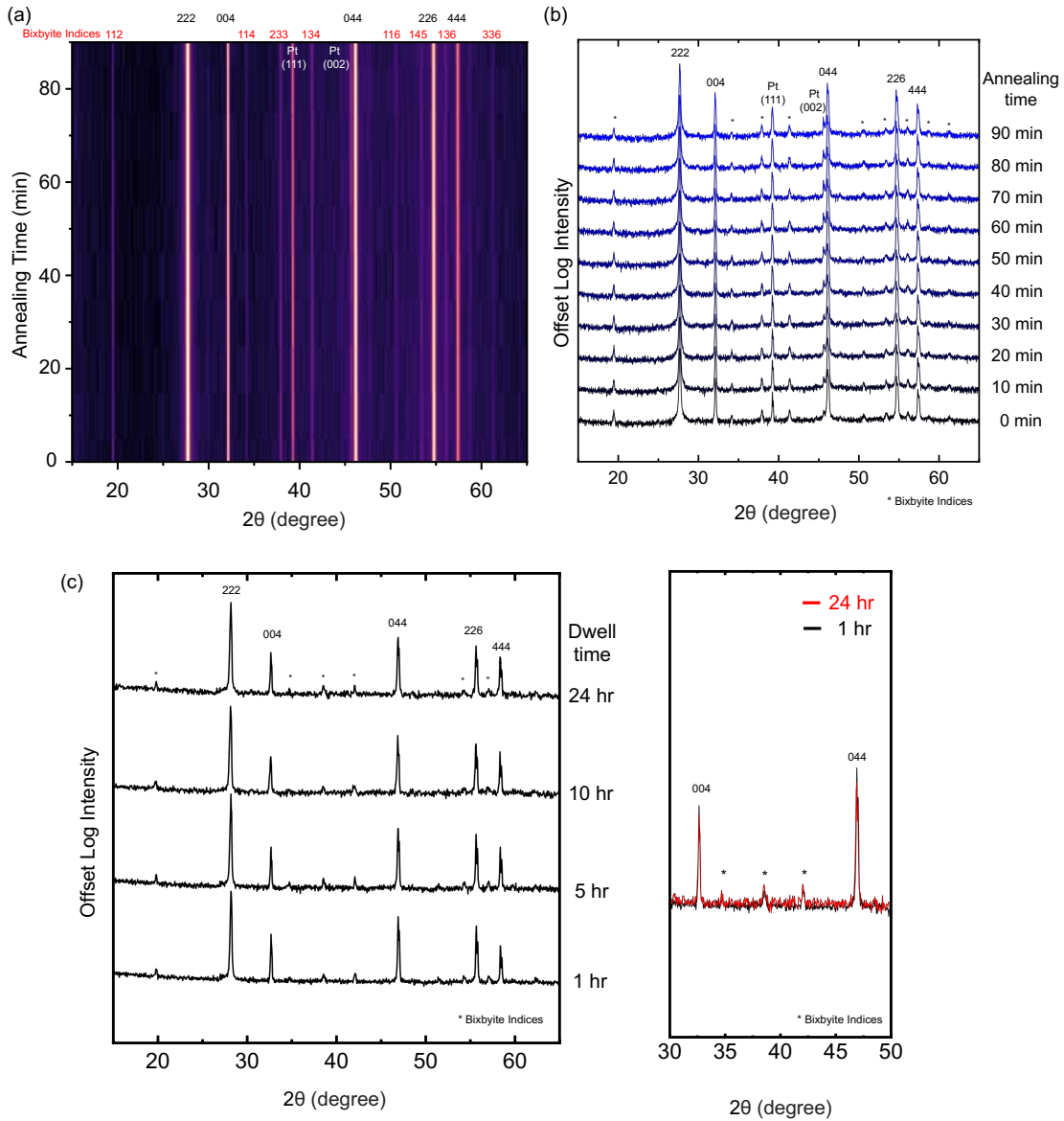

**Figure S2:** (a,b) time dependence in situ-HTXRD of $Ce_{0.325}(YLaPrSm)_{0.675}O_2$ showing phase stability under 1500 °C (c) ex situ XRD of $Ce_{0.325}(YLaPrSm)_{0.675}O_2$ under 1500 °C with different sintering time.

# Notes 3: Temperature dependence ex-situ X-ray diffraction patterns for constructing the phase diagram

Figure S3 summarizes the temperature dependence ex situ XRD patterns of $Ce_x(YLaPrSm)_{1-x}O_{2-\delta}$ (x=0.2, 0.35, 0.4, 0.5, 0.8) used for constructing the phase diagram in (**Figure 1**), where the grey arrows point to secondary phases.

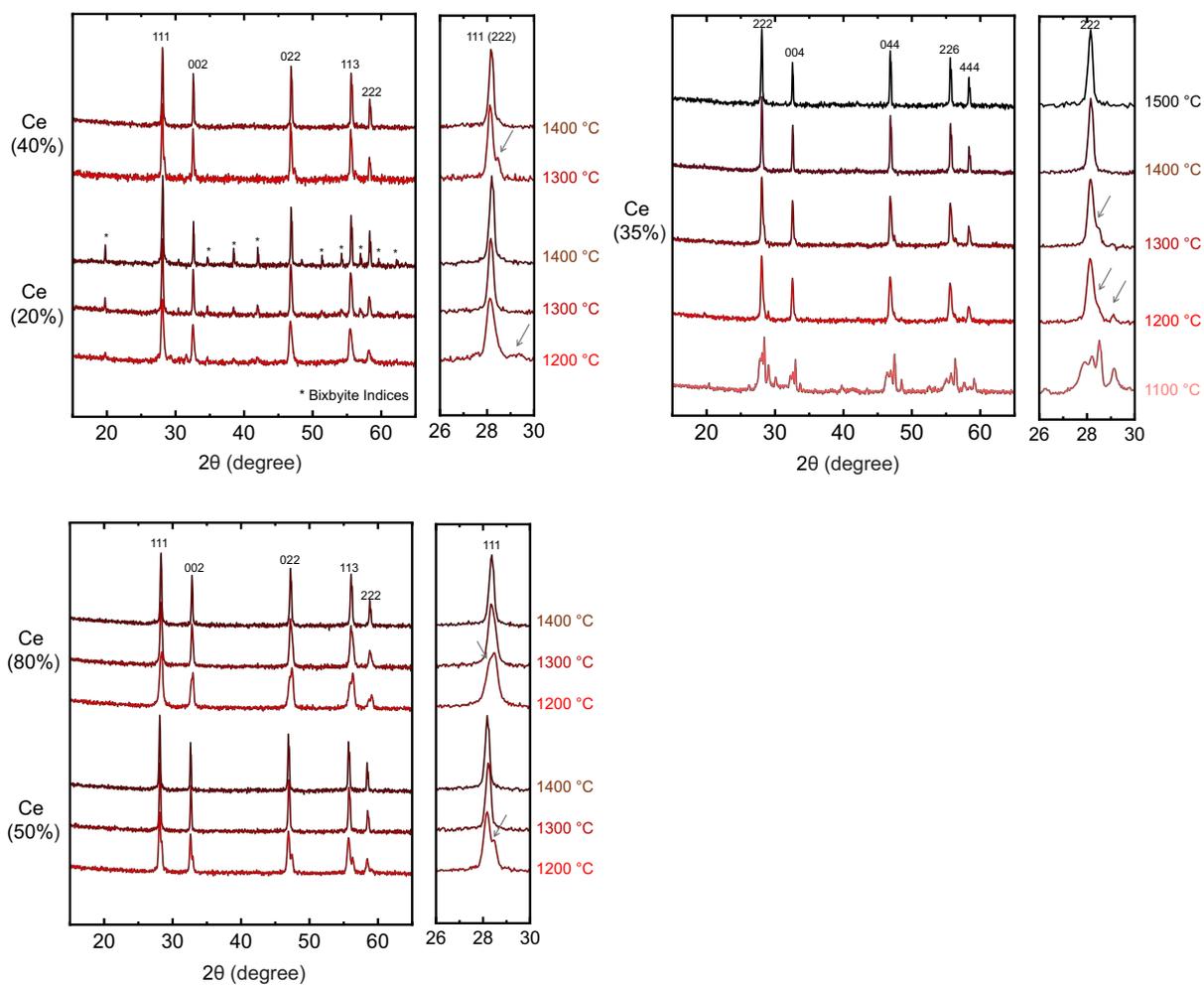

**Figure S3:** Temperature dependence ex situ XRD pattern of $Ce_x(YLaPrSm)_{1-x}O_{2-\delta}$ (x=0.2, 0.35, 0.4, 0.5, 0.8)

**Notes 4: SEM images and EDS mapping for the sintered $Ce_x(YLaPrSm)_{1-x}O_{2-\delta}$ pellets**

**Figure S4 (a-i)** shows the collected SEM images and the corresponding EDS maps for each synthesized pellet with different Ce concentrations corresponding to the 1500°C line in the phase diagram Figure 1(b) in the main manuscript. The pellets show similar grain sizes in the SEM, and the EDS maps show that all rare earth cations are homogeneously distributed with no detectable chemical segregation at the scales probed. The EDS spectra also show no surface contamination that could influence other characterization. **Figure S3 (j) and Table S1** summarize the calculated Ce concentration from EDS corresponding to the expected value. The error bar was calculated from measurements taken at five different locations on each pellet.

**Table S1:** Summary of the EDS quantification results taken at five different positions and their standard deviation

| Sample (Ce concentration) | Average Ce concentration (%) | Standard Deviation (%) |
| --- | --- | --- |
| 20 % | 20.0 | 0.19 |
| 30 % | 31.0 | 0.05 |
| 32.5 % | 32.2 | 0.07 |
| 35 % | 36.8 | 0.62 |
| 40 % | 41.5 | 0.21 |
| 50 % | 51.4 | 0.64 |
| 60 % | 62.0 | 0.35 |
| 70 % | 70.5 | 0.72 |
| 80 % | 79.9 | 0.78 |

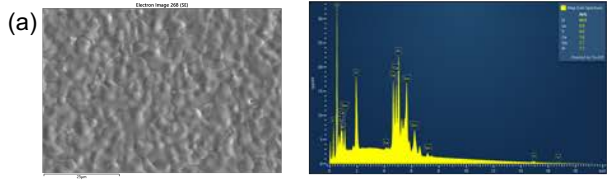
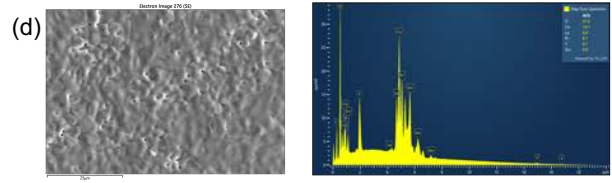
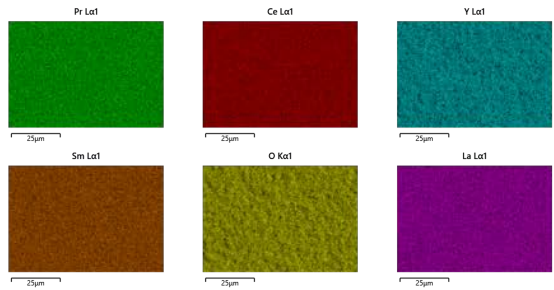
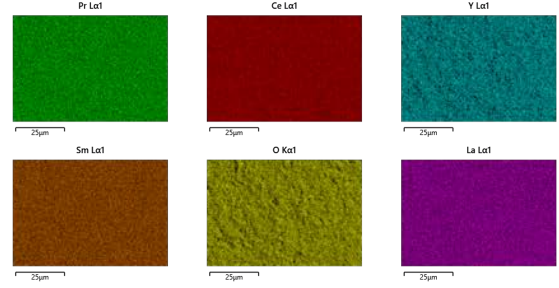
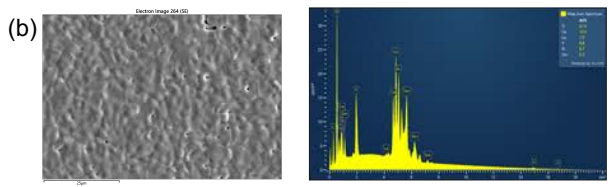
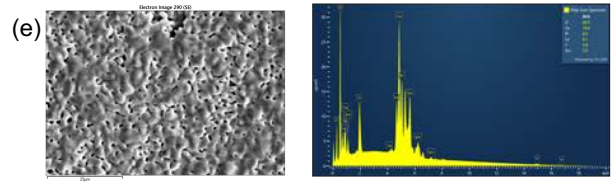
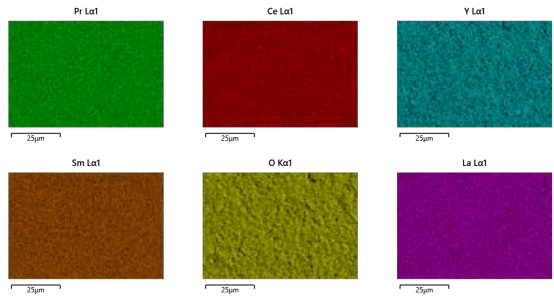
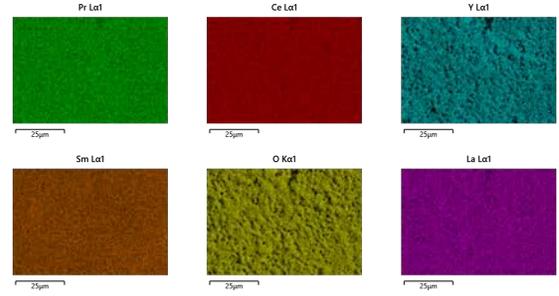
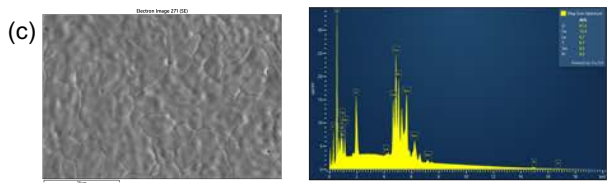
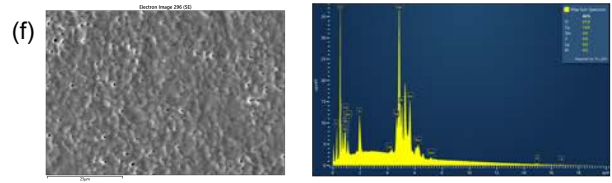
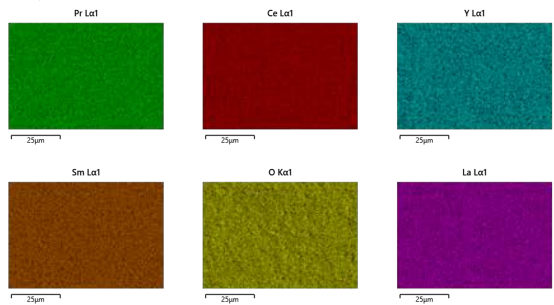
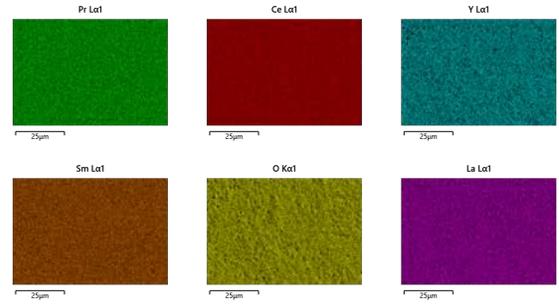

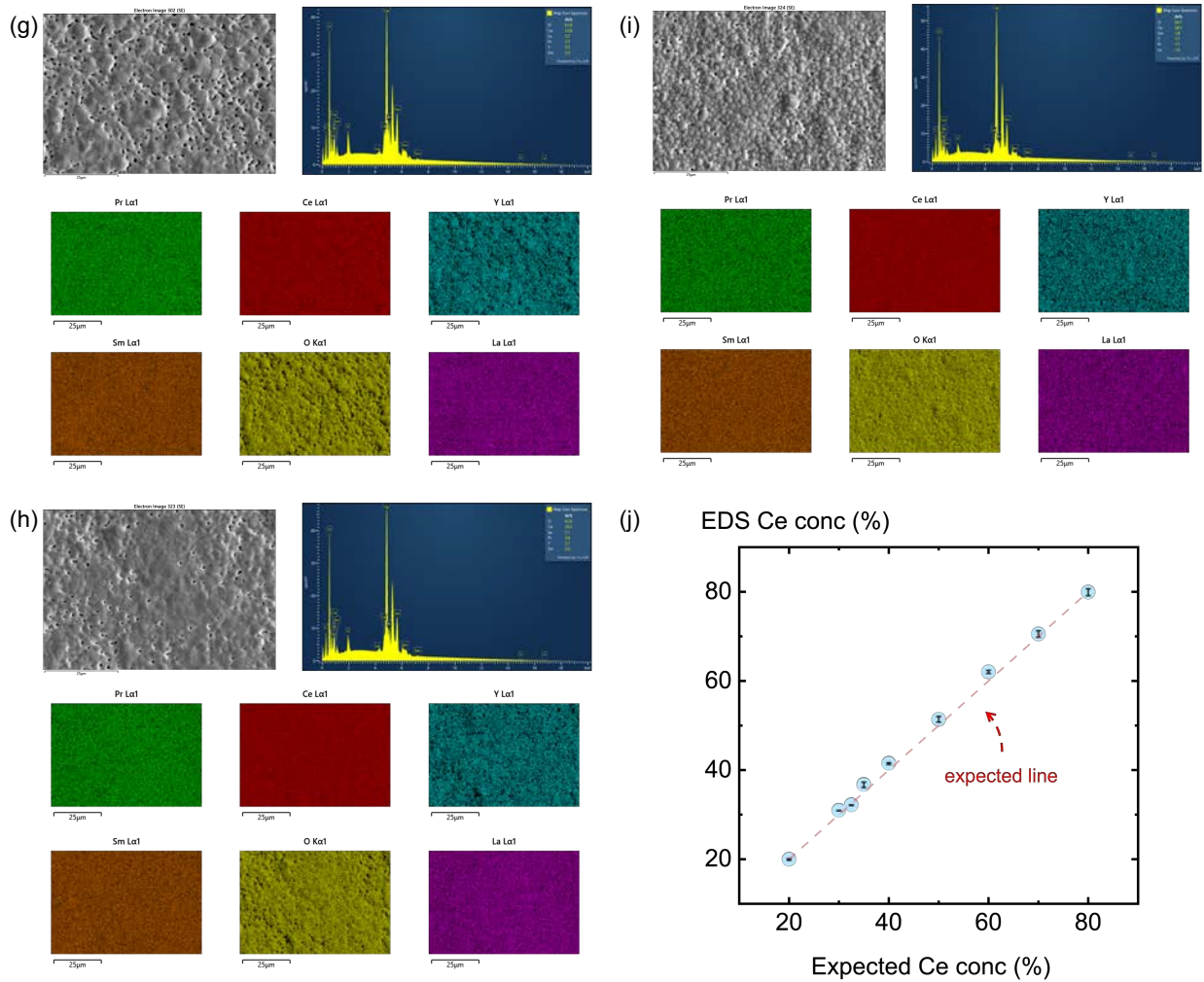

**Figure S4:** SEM and EDS mapping for $Ce_x(YLaPrSm)_{1-x}O_{2-\delta}$ (a) 20% (b) 30% (c) 32.5% (d) 35% (e) 40% (f) 50% (g) 60% (h) 70% (i) 80%, and (j) expected Ce concentration vs the EDS calculated concentration

**Notes 5: Full Range in situ high temperature X-ray diffraction pattern**

Figure S5 (a) 2D and (b) 3D plot of the full-range in situ high-temperature X-ray diffraction data for $Ce_{0.325}(YLaPrSm)_{0.675}O_{2-\delta}$, corresponding to Figure 1(e) in the main text.

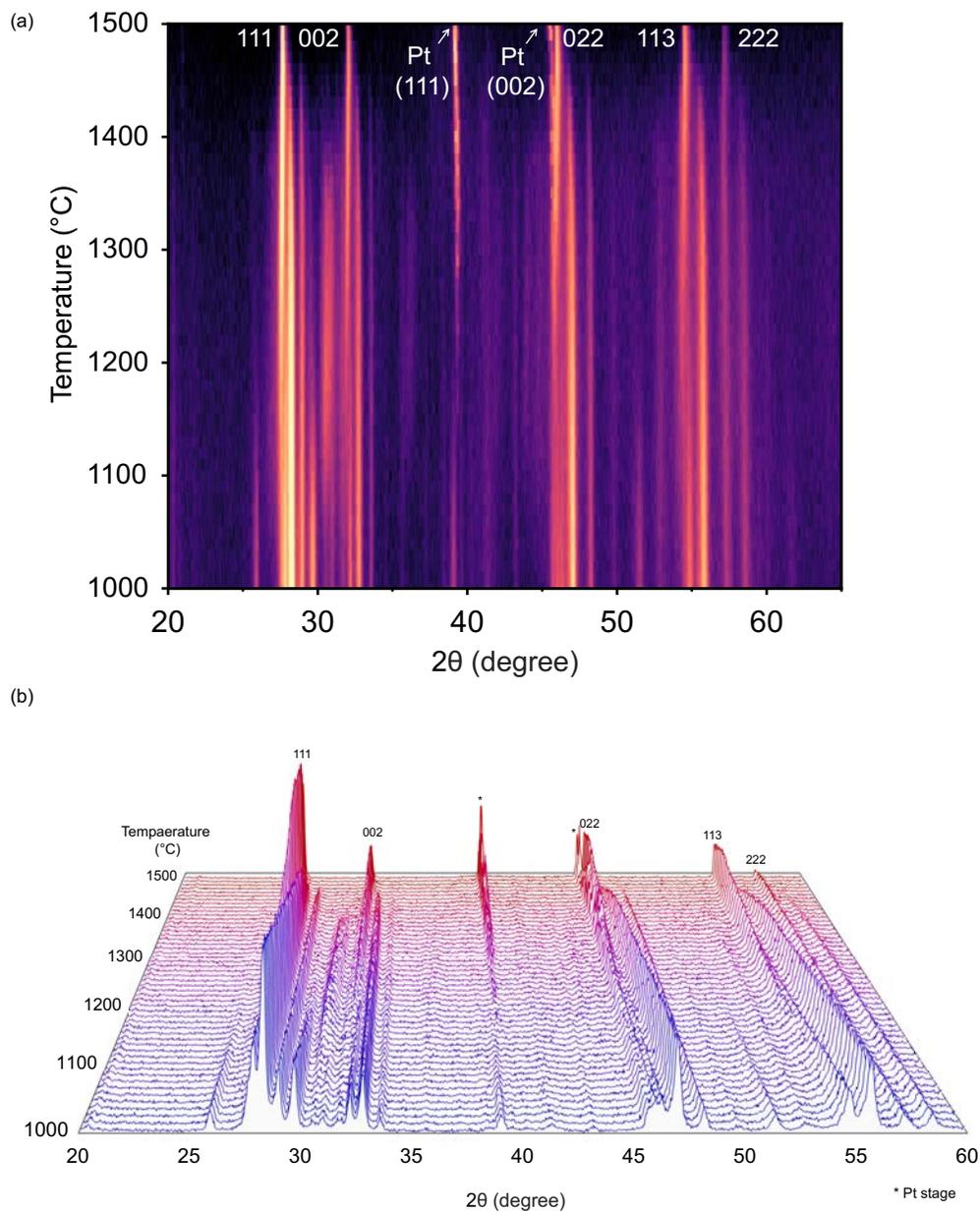

**Figure S5**: (a) 2D and (b)3D plot of the in situ high temperature XRD pattern of $Ce_{0.325}(YLaPrSm)_{0.675}O_{2-\delta}$

## Notes 6: Raman Shifts

Figure S6 shows the shifts of all Raman bands corresponding to the spectra in Figure 3, where the peak positions are simply determined by their highest intensity.

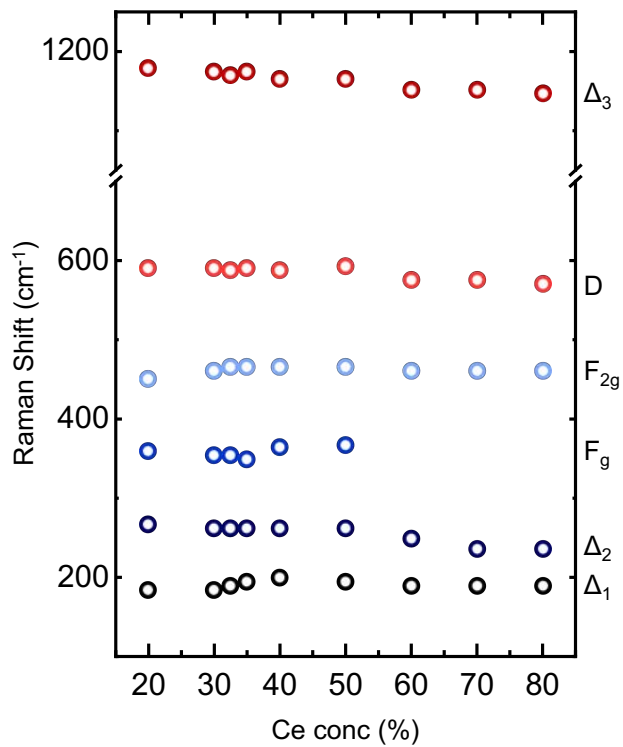

**Figure S6** Raman shift for each band in $Ce_x(YLaPrSm)_{1-x}O_{2-\delta}$

**Notes 7: Diffuse reflectance spectroscopy for confirming the optical band gaps for $Ce_x(YLaPrSm)_{1-x}O_{2-\delta}$ (x=0.2, 0.35, 0.4, 0.5, 0.8)**

Figure S7 shows the UV-vis spectra obtained from Diffuse reflectance spectroscopy, where the details of measurement are mentioned in the method section. The direct bandgap was determined using the Tauc relation and the Kubelka–Munk function; all Ce 20%, 30%, 32.5%, 40%, 50%, and 80% samples exhibit bandgap values around 2.05 eV.

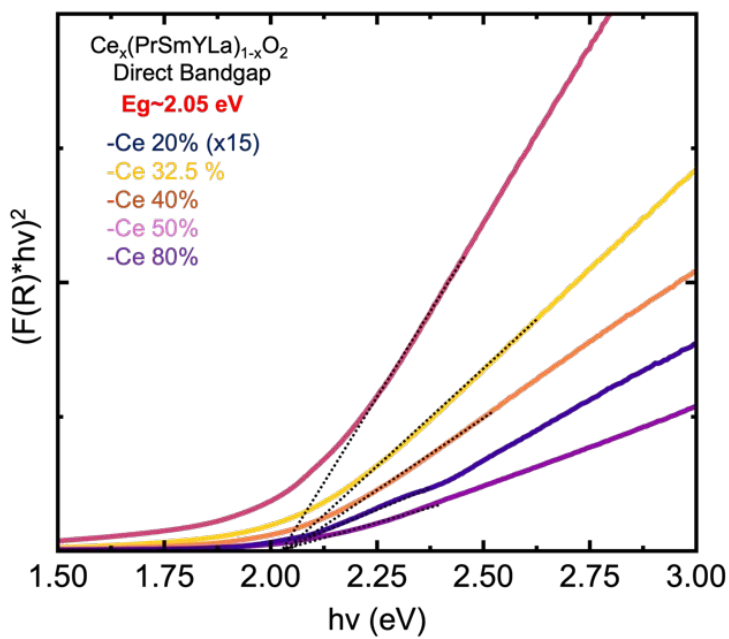

**Figure S7**: UV-Vis spectra for $Ce_x(YLaPrSm)_{1-x}O_{2-\delta}$ (x=0.2, 0.325, 0.4, 0.5, 0.8)

**Notes 8: The electronic structure analysis of the $Ce_x(YLaPrSm)_{1-x}O_{2-\delta}$ system via XANES**

Figure S8 summarizes the X-ray absorption near-edge structure (XANES) spectra of the lanthanide $L_3$ edges and the Y K edge for $Ce_x(YLaPrSm)_{1-x}O_{2-\delta}$ single-phase powders with x = 0.2, 0.325, and 0.4, each absorption edge spectra is compared to our corresponding parent oxide standard. Table S2 summarizes the edge energies extracted from Figure S8. This data is regrouped and reproduced from our XANES focused manuscript[13]: G. R. Bejger, M. K. Caucci, S. S. I. Almishal, B. Yang, J. Maria, S. B. Sinnott and C. M. Rost, *J. Mater. Chem. A*, 2025, 13, 29060 DOI: 10.1039/D5TA03815D.

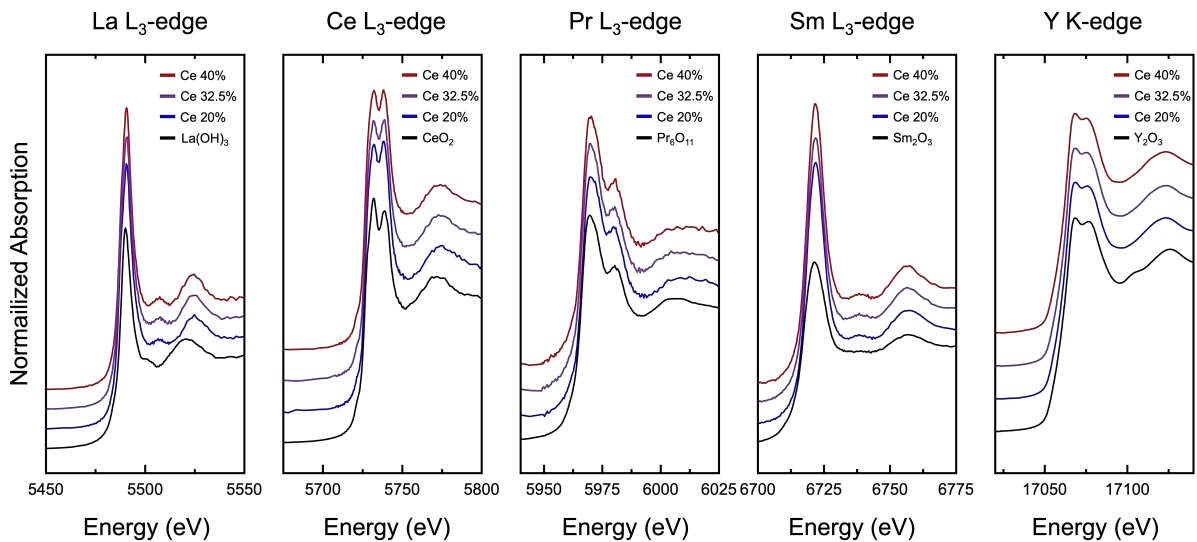

**Figure S8** X-ray absorption near edge structure (XANES) measurements of lanthanide $L_3$ edges and the Y K edge in $Ce_x(YLaPrSm)_{1-x}O_{2-\delta}$ (x=0.2, 0.325, and 0.4) powders and their parent oxides spectra

**Table S2:** $E_o$ values for the measured HEO series and standards. $E_o$ was selected to be the first peak in the first derivative of the absorption spectra.

|  | Ce $L_3$ (5723 eV) | Pr $L_3$ (5964 eV) | La $L_3$ (5483 eV) | Sm $L_3$ (6716 eV) | Y K (17038 eV) | |
|---|---|---|---|---|---|---|
| Ce 20% | 5726.59 | 5967.18 | 5486.92 | 6718.42 | 17054.98 | |
| Ce 32.5 % | 5726.39 | 5967.06 | 5486.82 | 6718.50 | 17055.00 | |
| Ce 40% | 5726.44 | 5966.83 | 5487.33 | 6718.67 | 17055.10 | |
| $CeO_2$ | 5726.42 | -- | -- | -- | -- | |
| $Pr_6O_{11}$ | -- | 5966.33 | -- | -- | -- | |
| $La_2(OH)_3$ | -- | -- | 5488.06 | -- | -- | |
| $Sm_2O_3$ | -- | -- | -- | 6717.32 | -- | |
| $Y_2O_3$ | -- | -- | -- | -- | 17055.01 | |

**Notes 9: Reversibility experiments on $Ce_{0.4}(YLaPrSm)_{0.6}O_{2-\delta}$ pellet**

Figure S9 shows the SEM images and the corresponding EDS maps for the pellet annealed under reducing conditions. The EDS results show that all rare-earth cations remain homogeneously distributed, and no surface contamination occurred during the annealing process that could contribute to the phase transition.

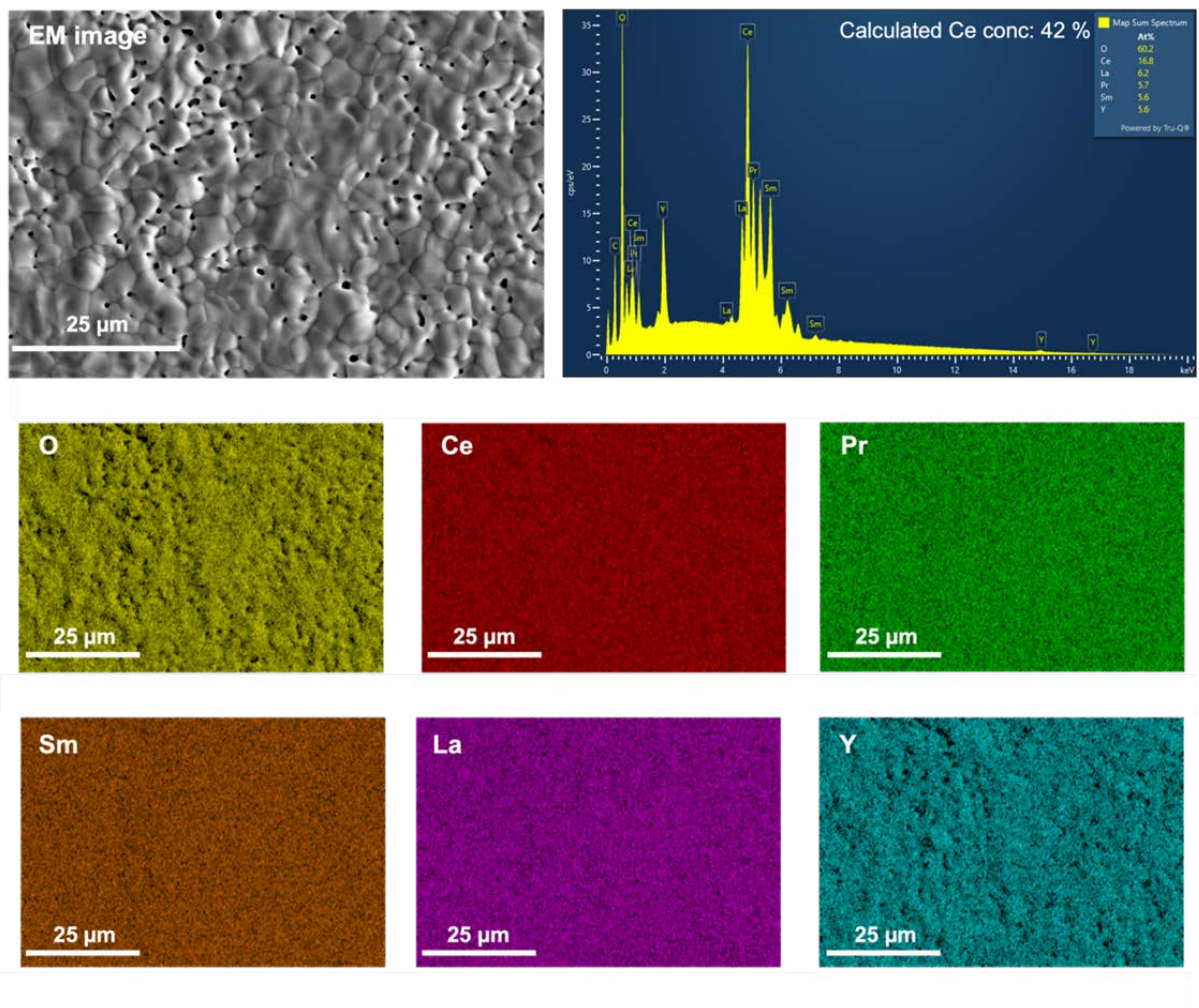

**Figure S9**: SEM and EDS mapping for the $Ce_{0.4}(YLaPrSm)_{0.6}O_{2-\delta}$ after annealing in the reducing environment

Figure S10 shows the supplementary XPS data for Figure 5, where Sm, Pr, La, Y, Ce 3d, and O 1s peaks are taken and processed as mentioned in the Method section. With the Ce oxidation-state change mentioned in the main text, Sm, Pr, La, and Y retain their oxidation states, with no chemical shift observed.

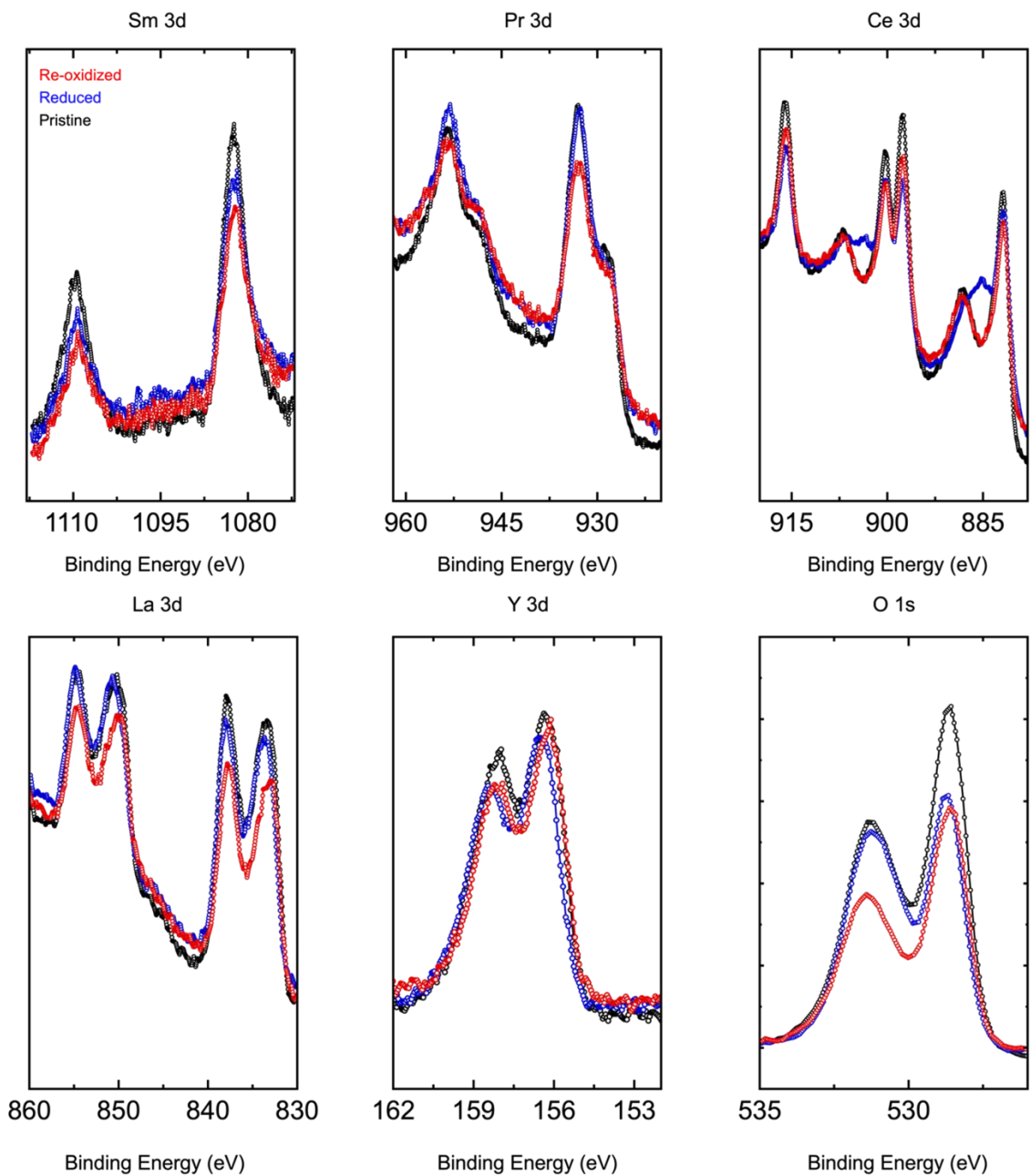

**Figure S10**: XPS spectra of the $Ce_{0.4}(YLaPrSm)_{0.6}O_{2-\delta}$ under different gas condition

**Notes 10: The thin film growth of CeYLaPrSO$_{2-\delta}$ via pulsed laser deposition**

Figure S11 shows the thin-film XRD data used to construct the growth map in Figure 6 in the main manuscript, with varying the laser fluence and growth temperature. Figure S12 shows that the thin film (sample #1) grown on a step and terrace YSZ substrate exhibits a smooth surface as probed by atomic force microscopy.

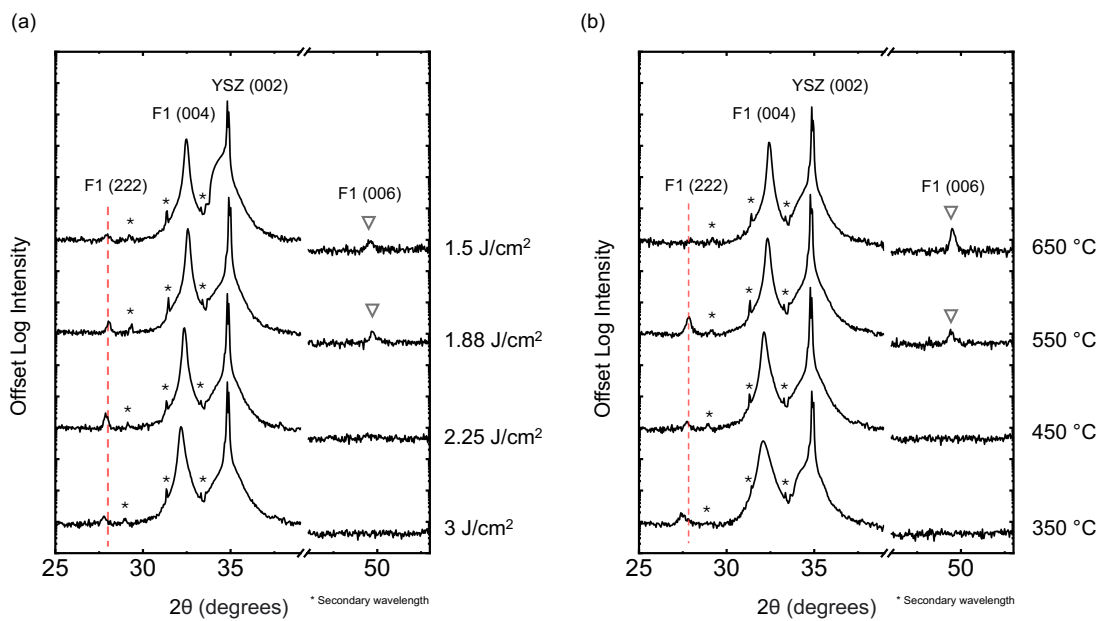

**Figure S11**: XRD pattern of CeYLaPrSO$_{2-\delta}$ under different growth conditions (a) different laser energy with fixing temperature at 600 degrees (b) different growth temperature with fixing laser energy at 120 mJ

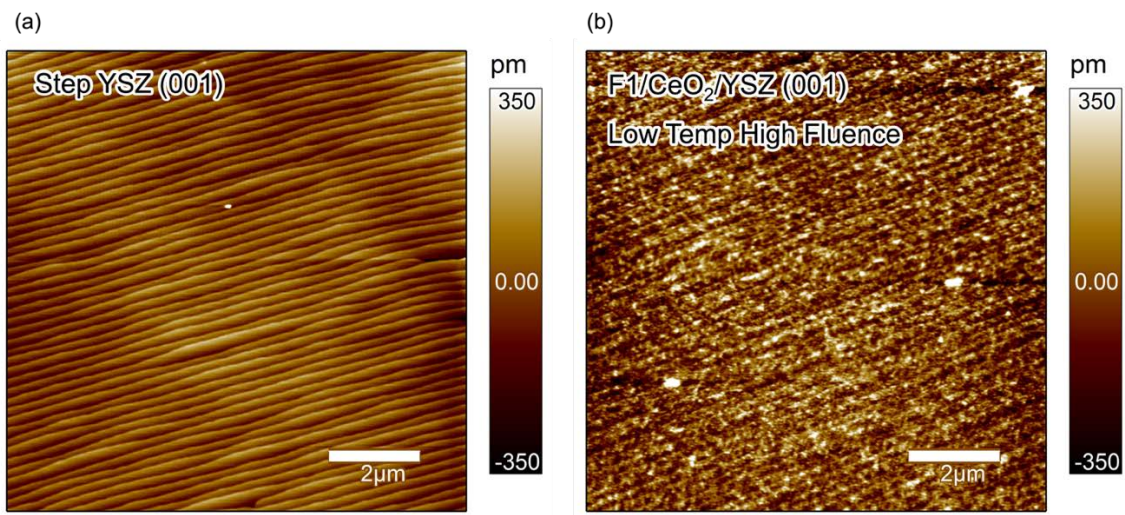

**Figure S12:** AFM tomography of (a) step YSZ (001) substrate (b) CeYLaPrSO$_{2-\delta}$ film with ceria buffer layer (sample #2)

Figure S13 a shows the HRTEM image of sample #1 with insets corresponding to the regions used to acquire the FFTs shown in panels **b** and **c**, respectively. The FFT of the HEO sample is consistent with the SAED pattern shown in Figure 6, confirming that the low-fluence, high-substrate-temperature film (sample #1) exhibits $I2_13$ symmetry. The EDS maps indicate homogeneous elemental distribution at this scale.

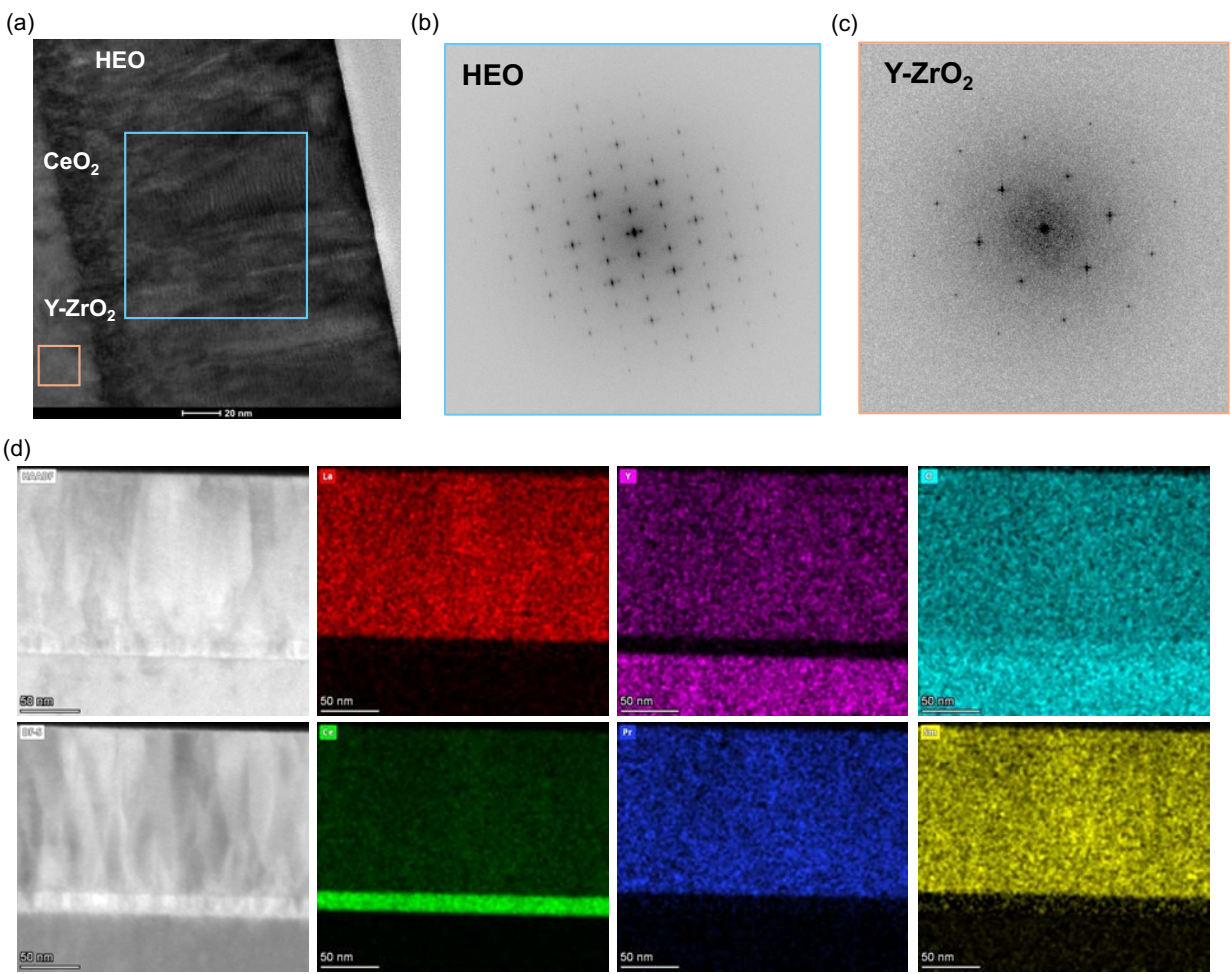

**Figure S13:** $Ce_{0.2}(YLaPrSm)_{0.8}O_{2-\delta}$ thin film (sample #1) (a) HR-TEM cross-section image of the thin film stack (b,c) FFT pattern of the selected thin film and substrate area along [110] direction (d) HAADF image and EDS mapping on the thin film

Figure S14 a shows the HRTEM image of sample #2 with insets corresponding to the regions used to acquire the FFTs shown in panels **b** and **c**, respectively. The FFT of the HEO sample is consistent with the SAED pattern shown in Figure 6, displaying certain missing reflections compared to fully ordered bixbyite symmetries, thereby confirming that the high-fluence, low-temperature film (sample #2) exhibits a "disordered fluorite symmetry". The EDS maps indicate a homogeneous elemental distribution at this scale.

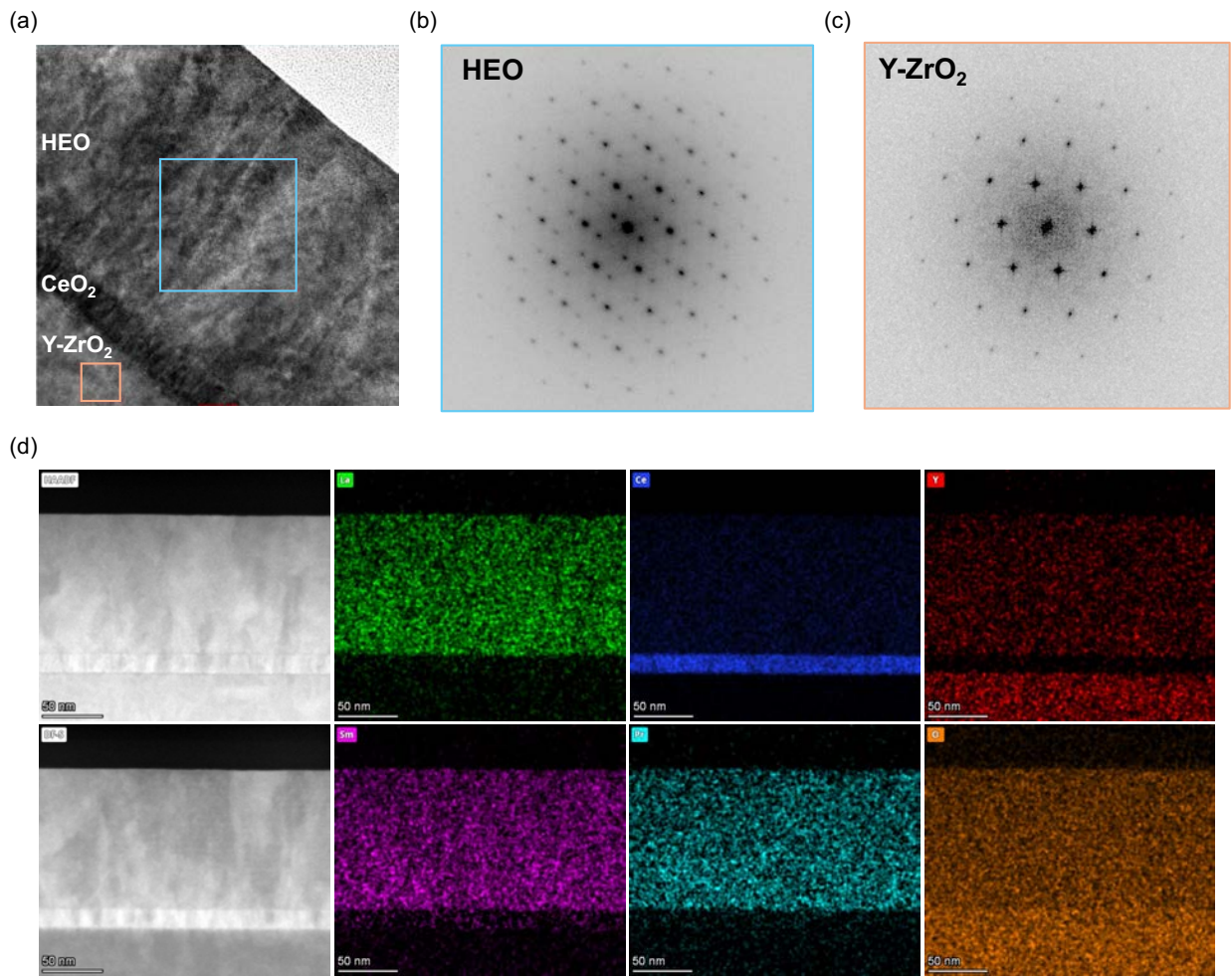

**Figure S14:** $Ce_{0.2}(YLaPrSm)_{0.8}O_{2-\delta}$ thin film (sample #2) (a) HR-TEM cross-section image of the thin film stack (b,c) FFT pattern of the selected thin film and substrate area along [110] direction (d) HAADF image and EDS mapping on the thin film

Figure S15 a shows the HRTEM image of sample #3 with insets corresponding to the regions used to acquire the FFTs shown in panels **b** and **c**, respectively. The FFT of the HEO sample closely resembles the SAED pattern shown in Figure 6d, displaying certain missing reflections compared to fully ordered bixbyite symmetries, thereby confirming that the high-fluence, low-temperature film without ceria buffer sample #3 also exhibits a "disordered fluorite symmetry". The EDS maps indicate a homogeneous elemental distribution at this scale.

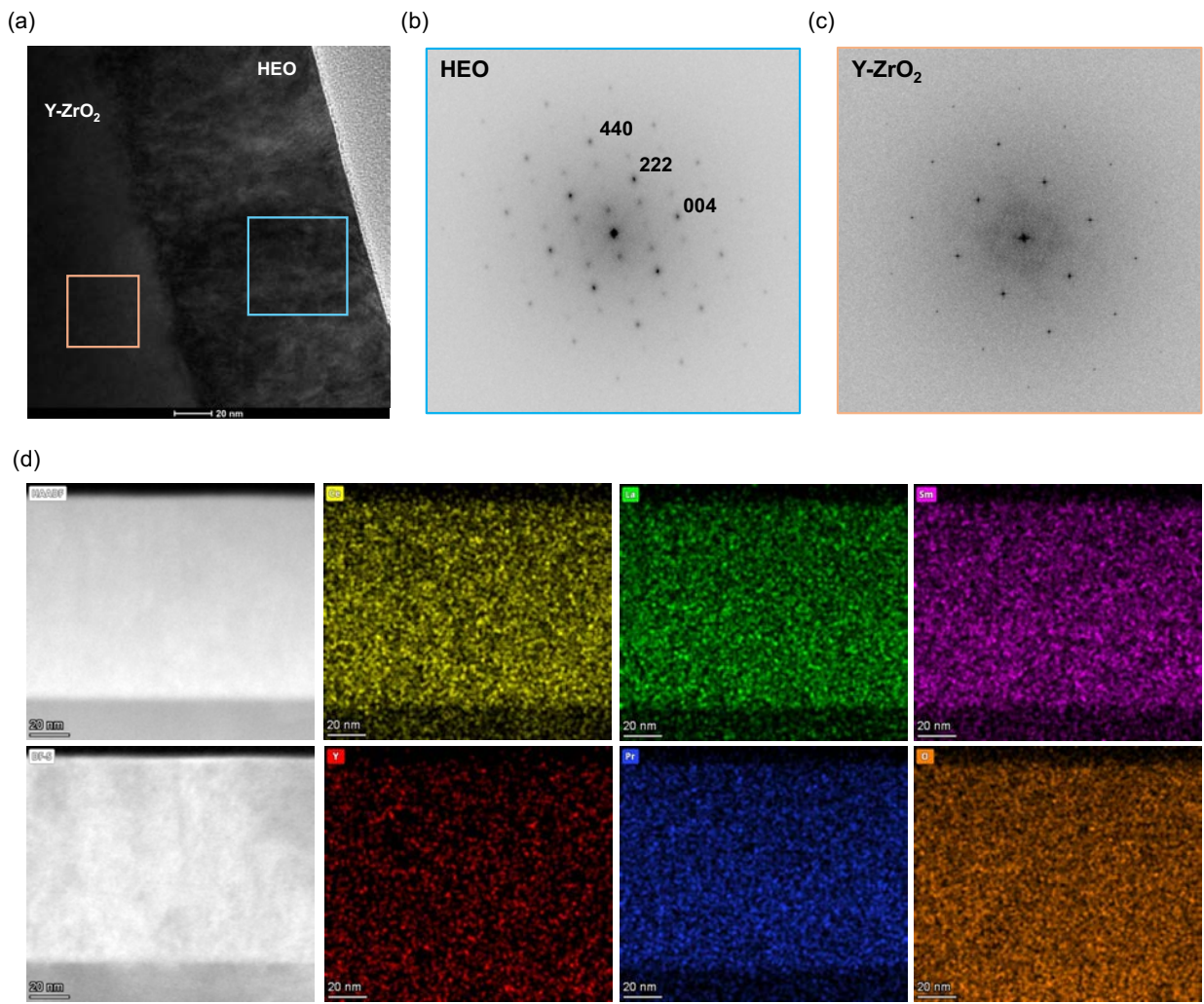

**Figure S15:** $Ce_{0.2}(YLaPrSm)_{0.8}O_{2-\delta}$ thin film (sample #3) HR-TEM cross-section image of the thin film stack (b,c) FFT pattern of the selected thin film and substrate area along [110] direction (d) HAADF image and EDS mapping on the thin film

## Note 11: Details of the electrochemical impedance measurements

Table S3 summarizes all details and results for the conductivity measurements using electrochemical impedance spectroscopy. Figure S16 (a,b) shows the temperature and Ce concentration dependent conductivity values determined by the lowest Z'' point from the Nyquist plot S16 (c-f) with reference values of 10 % doped GDC, 20% doped GDC, and YSZ inserted[33,34].

**Table S3:** Summary of the experimental details for the impedance measurements and the calculated conductivity values at temperatures from 400 to 700 °C

| Sample | Thickness (mm) | Electrode diameter | Activation Energy (eV) | Linear Fit $R^2$ value (%) |
|---|---|---|---|---|
| Ce 20% | 1.18 | 8 mm | 0.60 | 99.99 |
| Ce 32.5% | 1.23 | | 0.60 | 99.98 |
| Ce 50% | 1.07 | | 0.84 | 99.83 |
| Ce 80% | 1.20 | | 0.79 | 99.86 |

| Sample | Temperature (°C) | $R_{total}$ (ohm) | Conductivity (S cm$^{-1}$) |
|---|---|---|---|
| Ce 20% | 400 | 497 | $4.73 \cdot 10^{-4}$ |
| | 500 | 156 | $1.51 \cdot 10^{-3}$ |
| | 600 | 61.5 | $3.82 \cdot 10^{-3}$ |
| | 700 | 30.2 | $7.79 \cdot 10^{-3}$ |
| Ce 32.5% | 400 | 826 | $2.96 \cdot 10^{-4}$ |
| | 500 | 255 | $9.59 \cdot 10^{-4}$ |
| | 600 | 97.5 | $2.51 \cdot 10^{-3}$ |
| | 700 | 49.1 | $4.99 \cdot 10^{-3}$ |
| Ce 50% | 400 | 1561 | $1.36 \cdot 10^{-4}$ |
| | 500 | 276.3 | $7.71 \cdot 10^{-4}$ |
| | 600 | 65.2 | $3.27 \cdot 10^{-3}$ |
| | 700 | 27.9 | $7.64 \cdot 10^{-3}$ |
| Ce 80% | 400 | 776 | $3.08 \cdot 10^{-4}$ |
| | 500 | 136 | $1.76 \cdot 10^{-3}$ |
| | 600 | 39.9 | $5.97 \cdot 10^{-3}$ |
| | 700 | 17.0 | $1.41 \cdot 10^{-2}$ |

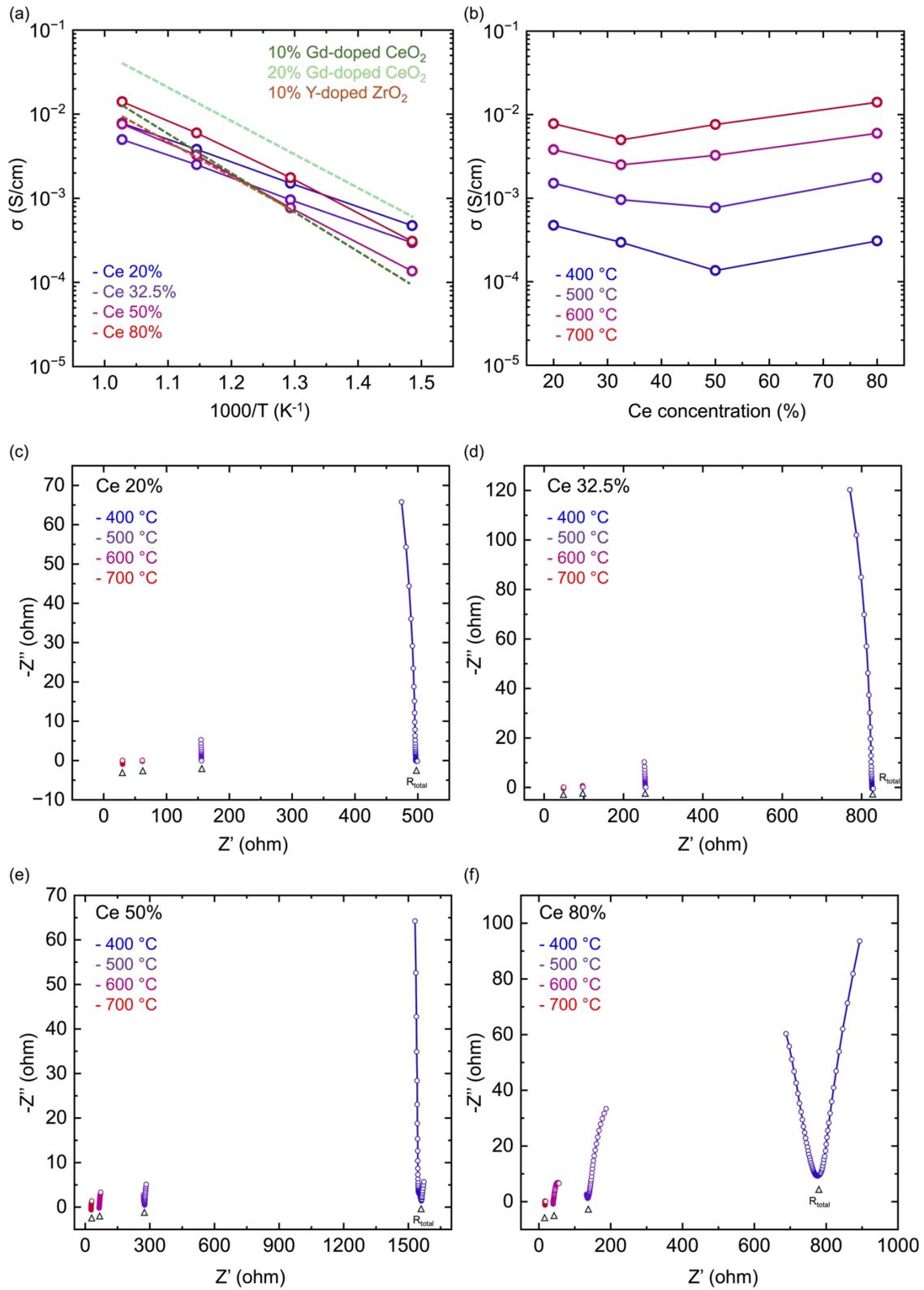

**Figure S16**: (a) Conductivity corresponds to Ce concentration with reference conductivity value of 10 % Gd doped $CeO_2$, 20 % Gd doped $CeO_2$, and 10 % Y doped $ZrO_2$ (b) conductivity correapond to temperature (c,d,e,f) Nyquist plot of the pellets with composition of $Ce_x(YLaPrSm)_{1-x}O_{2-\delta}$ (x=0.2, 0.325, 0.5, 0.8)